\documentclass[journal=jctcce,manuscript=article,layout=twocolumn]{achemso}

\usepackage[version=3]{mhchem} 
\usepackage[usenames]{color}
\usepackage[normalem]{ulem}
\usepackage{braket}
\usepackage{subcaption}
\usepackage{graphicx}
\usepackage{amssymb}

\newcommand{\lan}{\langle}
\newcommand{\ran}{\rangle}

\newcommand{\affpcl}{Parallel Computing Lab, Intel Corporation, Santa Clara, CA 95054, USA}
\newcommand{\affccb}{Department of Chemistry and Chemical Biology, Harvard University, Cambridge, MA 02138, USA}
\newcommand{\afflbl}{Computational Research Division, Lawrence Berkeley National Laboratory, Berkeley, CA 94720, USA}
\newcommand{\shortpcl}{Parallel Computing Lab, Intel Corporation}
\newcommand{\shortccb}{Department of Chemistry and Chemical Biology, Harvard University}
\newcommand{\shortlbl}{Computational Research Division, Lawrence Berkeley National Laboratory}

\newcommand{\qhipster}{\textit{q}H\textit{i}PSTER}

\author{Nicolas P. D. Sawaya}
\affiliation[\affccb]{\shortccb}
\alsoaffiliation[\affpcl]{\shortpcl}

\author{Mikhail Smelyanskiy}
\affiliation[\affpcl]{\shortpcl}

\author{Jarrod R. McClean}
\affiliation[\afflbl]{\shortlbl}

\author{Al\'{a}n Aspuru-Guzik}
\affiliation[\affccb]{\shortccb}
\email{aspuru@chemistry.harvard.edu}

\title{Error Sensitivity to Environmental Noise in Quantum Circuits for Chemical State Preparation.}


\begin{document}

\begin{tocentry}

\includegraphics[width=8.9cm]{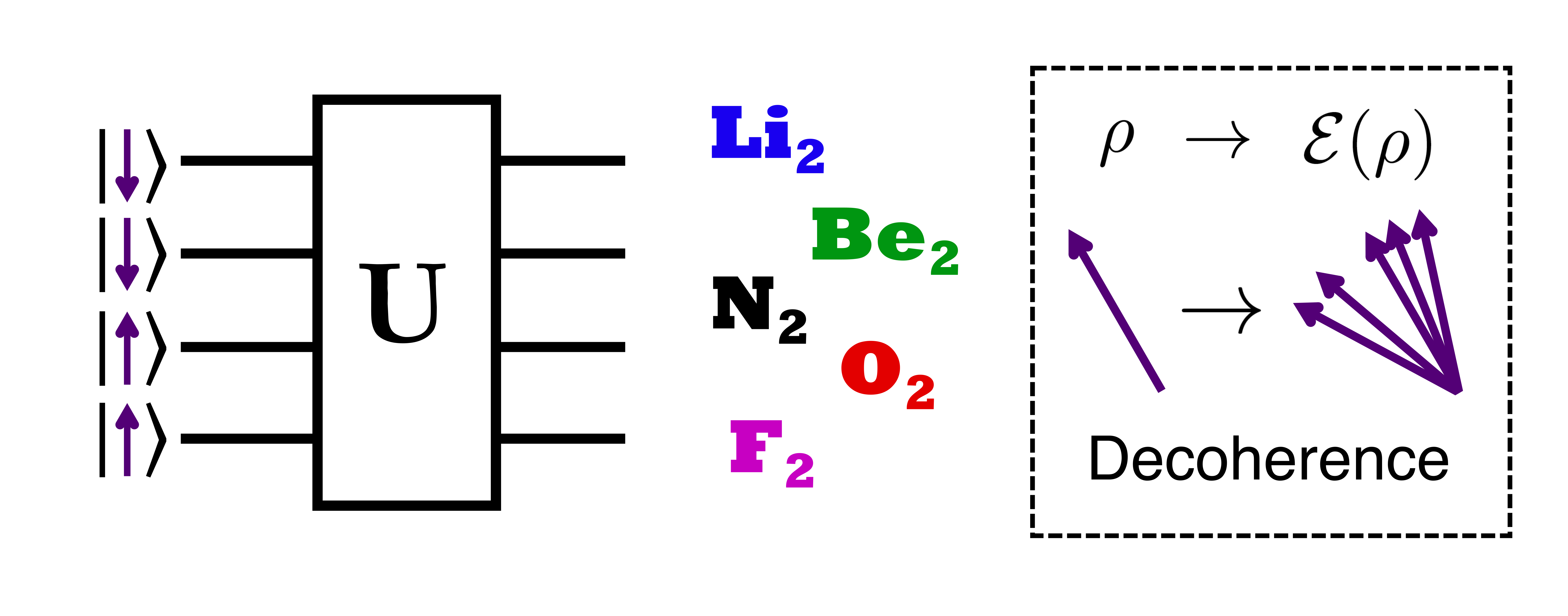}





\end{tocentry}



\begin{abstract}

Calculating molecular energies is likely to be one of the first useful applications to achieve quantum supremacy, performing faster on a quantum than a classical computer. However, if future quantum devices are to produce accurate calculations, errors due to environmental noise and algorithmic approximations need to be characterized and reduced. In this study, we use the high performance \textit{q}H\textit{i}PSTER software to investigate the effects of environmental noise on the preparation of quantum chemistry states. We simulated eighteen 16-qubit quantum circuits under environmental noise, each corresponding to a unitary coupled cluster state preparation of a different molecule or molecular configuration. Additionally, we analyze the nature of simple gate errors in noise-free circuits of up to 40 qubits. We find that, in most cases, the Jordan--Wigner (JW) encoding produces smaller errors under a noisy environment as compared to the Bravyi--Kitaev (BK) encoding. For the JW encoding, pure-dephasing noise is shown to produce substantially smaller errors than pure relaxation noise of the same magnitude. We report error trends in both molecular energy and electron particle number within a unitary coupled cluster state preparation scheme, against changes in nuclear charge, bond length, number of electrons, noise types, and noise magnitude. These trends may prove to be useful in making algorithmic and hardware-related choices for quantum simulation of molecular energies.

\end{abstract}


\section{Introduction}
\label{sec:intro}

For certain classes of problems, quantum computation promises to provide polynomial or exponential speedups compared to classical computers \citep{montanaro2015,MikeIke}. Despite the rapid progress in obtaining longer qubit coherence times \citep{martinisEcc2015,Senko2015,Schindler2013}, current experimental efforts to demonstrate quantum circuits employ pre-threshold quantum information processors \citep{MikeIke,RiefPolak,OmalBab15}. Hence it is essential to determine the effects that noise has both on error-correction codes and on specific algorithms. By characterizing and understanding error trends, one can determine which aspects of a particular quantum device or algorithm need to be improved.

Here, we use the recently developed \qhipster~software \citep{qhipster} to simulate quantum state preparation circuits for the molecular electronic structure problem. We use circuits that prepare unitary coupled cluster states of a set of seven diatomic molecules, as well as multiple bond lengths along the dissociation coordinate of the nitrogen molecule. We also study and compare three isoelectronic diatomic species. Finally, we simulate short chemistry-inspired circuits on up to 40 qubits, approximating single-electron excitations from a reference state. In order to determine the effects of decoherence, the molecular quantum circuits are simulated under the presence of environmental noise. Studying the results reveals trends in error rates for molecular energy and electron particle number, which may help guide future algorithmic and hardware developments.

Two classes of quantum algorithms have emerged for calculating chemical energies. The first uses the quantum phase estimation (QPE) algorithm \citep{aag2005,abrams1999,veispittner2010,whitfield2011,reiher2016} to evolve an initial state toward the molecular ground state. The second approach is the variational quantum eigensolver (VQE) \citep{peruzzomcclean2014,mcclean2015,wecker2015_vqe}. Similar ideas to the VQE have been independently introduced and experimentally realized within the context of matrix product states \citep{barrett2013,eichler2015}.

In VQE, a quantum circuit first prepares a parametrized approximation to the quantum state, which is then used to estimate the energy by direct measurement of the expectation value of the terms of the Hamiltonian. Then, the variationally-optimal state is found by modifying the state parametrization and corresponding preparation circuit after each energy calculation, to find the state(s) which minimize the energy. For chemistry applications, VQE generally requires fewer coherent gate operations per circuit than QPE does, making the VQE approach a more promising candidate for near-term quantum hardware.

In the current study, we focus on the preparation of molecular states using the unitary coupled cluster (UCC) ansatz \citep{mcclean2015,ucc2006,ucc2007}. Quantum chemical state preparation is important in both the VQE approach, where it ultimately determines the state, energy, and properties measured, and in QPE, where it determines the success probability of the energy measurement \citep{aag2005,Wang:2008,McClean2014b}. Hence an understanding of the general effect of errors on state preparation is crucial to the success of both algorithms.

In addition to the choice between QPE and VQE, there are two common methods for mapping the electronic structure problem to a set of qubits: the Jordan--Wigner \citep{jw1928,whitfield2011} (JW) and Bravyi--Kitaev \citep{bk2002,love2012} (BK) transformations. It has been shown for the hydrogen molecule that the BK mapping requires fewer gate operations than JW, in order to achieve a given chemical precision \citep{love2012}. Moreover, the reduced locality of the BK transformed Hamiltonian has many benefits for adiabatic schemes and state preparation methods.  Each of these methods seems to display practical benefits and drawbacks that lend them best to different situations.  These and other studies have investigated gate ordering, gate cancellation, and the effects of various Trotterization approximations \citep{Hastings2015,babbush2015}.

The article is organized as follows. In Sec.~\ref{sec:bg} we present the physical background and methodology used in the study, including the noise model and the method for mapping the molecular electronic structure problem onto a set of qubits. In Sec.~\ref{sec:enrg} we study the effects of environmental noise on the molecular energy of the prepared state. Sec.~\ref{sec:enum} presents results for the effects of both noise and gate errors on electron number preservation. Finally, Sec.~\ref{sec:hpc} summarizes the implementation and performance details of \qhipster, before concluding remarks are given in Sec.~\ref{sec:concl}.

\section{Background and Methods}
\label{sec:bg}

\subsection{Environmental Noise in Quantum Circuits}

Various noise models have been developed to study environmental effects and general errors in noisy quantum systems and nonideal quantum circuits, varying in accuracy and computational complexity. In this section, we briefly introduce the models of noise relevant to our study. These methods can be roughly split into two categories: the direct use of the Schr\"{o}dinger or quantum master equations, and the use of heuristic error operators.

In the first category, time is explicitly considered and the Schr\"{o}dinger equation is used to propagate the state of the system. Highly accurate open quantum systems methods such as the quasiadiabatic propagator path integral method (QUAPI) \citep{quapi1,quapi2} are too computationally expensive to be used on more than a few qubits.  Commonly used master equations such as the Redfield formalism \citep{redfield1965} propagate the density matrix directly. Finally, state vector methods include the quantum jump formalism \citep{Plenio98} and various forms of the stochastic Schr\"{o}dinger equation \citep{Fox1978}. All of these methods often make use of a detailed bath spectral density, which describes the frequency-dependent system-bath interactions.

The second noise simulation category involves the use of one or more heuristic parameters that encapsulates the relevant noise effects. This strategy is used when the bath spectral density is unknown or its precise form is unimportant. Especially common in the simulation of quantum circuits for quantum computation, this simulation category is chosen when the goal is to elucidate general trends relating to a specific algorithm in its abstract circuit representation, as opposed to faithfully reproducing the results of a specific hardware setup. The operator-sum representation is often used for this type of noise modeling, in which the density matrix $\rho$ evolves as

\begin{equation}
\begin{aligned}
\rho \rightarrow \mathcal{E}(\rho) = \sum_i E_i \rho E_i^\dagger
\label{eq:opsumrep}
\end{aligned}	
\end{equation}

where ${E_i}$ are a set of Kraus operators. One example of such a quantum operation is the asymmetric depolarizing channel, defined by 

\begin{equation}
\begin{aligned}
\rho \rightarrow (1-p_x-p_y-p_z)\rho \\
+ p_x X \rho X + p_y Y \rho Y + p_z Z \rho Z
\label{eq:assymdepol}
\end{aligned}
\end{equation}

where $X$, $Y$, and $Z$ are the 
Pauli matrices, and the $p_{\{x,y,z\}}$ correspond to the probabilities of a Pauli operator acting on the system during one iteration of the channel. When multiple qubits are considered, Pauli matrices for each qubit are applied independently according to the formula above.

Other noise channels include the symmetric depolarizing, bit flip, phase flip, amplitude damping, and phase damping channels. The limits of two common noise channels are instructive: the symmetric depolarizing channel converges to a completely mixed state, and the amplitude damping channel causes decay toward the ground state. We note that, though we do not do so in this study, it is also possible include the effects of correlated noise \citep{preskill2006,Monz11}, which can lead to a faster total decay known as superdecoherence.

Determining noise-induced behavior is especially important if a physical implementation of a quantum circuit is to use gates--such as the precise Z rotations in our circuits--which require careful decomposition in an error correction scheme. The trade-off between imperfect arbitrary rotations and error correction schemes, which can be costly both in terms of gates and qubits, needs to be carefully studied in near-term devices where one may have sufficient resources to provide some, but not perfect, error correction \citep{solovaykitaev}. Additionally, we note that studying noise effects is important even if an implementation were to use error correction, because there will also be an effective decoherence time associated the error-corrected circuit, which may not be infinite on near-term devices.

\subsection{Pauli Twirling Approximation}

One of the challenging aspects of modeling environmental noise in quantum systems is that the effects become more dramatic as the system size increases.  As a result, the eventual aim is to be able to treat the largest number of qubits feasible with given computational resources. In order to scale up our simulations, it is preferable to propagate a state vector instead of the density matrix, because the state vector contains $2^n$ complex coefficients while the density matrix requires $2^{2n}$ terms, where $n$ is the number of qubits.  This is the strategy we employ in this work, and describe in more detail in this section.

It is possible to map any Lindblad-type master equation onto a linear stochastic differential equation (SDE) operating on the state vector \citep{jacobs1998,bassi2003,bassi08}. With such a mapping, many stochastic trajectories of the state vector are summed in order to converge to the same physical results as if propagating the full density matrix \citep{bassi08}. More specifically, here we use the Pauli twirling approximation (PTA), in which a noise channel is mapped onto a set of Pauli operators. Specifically, we use PTA to approximate decoherence by the dephasing and amplitude damping channels. This noise model approximation has been previously used in simulating error-correcting circuits \citep{silva2008,geller2013,tomita2014}, with the assumption that the circuit time is much shorter than the coherence time. This assumption is allowable since once the circuit time nears the coherence time, the algorithm has likely long ceased to be useful. In order to use PTA for propagating a state vector with a SDE, one draws $\nu_X$, $\nu_Y$, and $\nu_Z$ from three independent Gaussian distributions, and performs three ``noise gates'' in sequence, resulting in a unitary operator

\begin{equation}
\begin{aligned}
U_{noise} = \exp(-i\nu_X X) \exp(-i\nu_Y Y) \\ \times \exp(-i\nu_Z Z)
\label{eq:rot3}
\end{aligned}
\end{equation}

based on the sampled values \citep{bassi08}.

For small values of $\nu$, these three rotations can be approximated as a single rotation:
\begin{equation}
\begin{aligned}
U_{noise} = R_{\vec{n}}(\theta) \approx \exp(-i\frac{\theta}{2} {\vec{n}} \cdot \vec{\sigma})
\label{eq:rot1}
\end{aligned}
\end{equation}

 where
 
\begin{equation}
\begin{aligned}
\vec{n}=\frac{\vec{\nu}}{||\vec{\nu}||}
\label{eq:defn}
\end{aligned}
\end{equation}

\begin{equation}
\begin{aligned}
\theta = 2 ||\vec{\nu}||
\label{eq:deftheta}
\end{aligned}
\end{equation}

Independent operators $R_{\vec{n}}(\theta)$ are applied to each qubit after every time step. For a particular qubit, long strings of these noise gates may be fused together for time steps during which there is no logic gate performed on the said qubit. PTA converges with fewer iterations than performing rare probabilistic bit-flips and phase-flips on the state vector, which is one common algorithm used to model noise channels through state vector propagation \citep{DBLP:phd/de/Trieu2010}.

The Gaussian distributions from which $\nu_X$, $\nu_Y$, and $\nu_Z$ are drawn have $0$ mean and standard deviations $s_x$, $s_y$, and $s_z$ defined by

\begin{equation}
\begin{aligned}
s_\alpha = \sqrt{-\ln(1-p_\alpha)} 
\label{eq:def_s}
\end{aligned}
\end{equation}

with $\alpha \in \{x,y,z\}$ and
\begin{equation}
\begin{aligned}
p_x = p_y = \frac{1-e^{-1/M_1}}{4}
\label{eq:defpxpy}
\end{aligned}
\end{equation}

\begin{equation}
p_z = \frac{1-e^{-1/M_2}}{2} - \frac{1-e^{-1/M_1}}{4}
\label{eq:defpz}
\end{equation}

where we denote $M_\beta=T_\beta/\Delta t$ the ``coherence parameter'' and $\beta \in \{1,2,\phi\}$. The time $T_\phi$ is related to the standard $T_1$ and $T_2$ times as 

\begin{equation}
\frac{1}{T_\phi} = \frac{1}{T_2} - \frac{1}{2T_1}
\label{eq:defT}
\end{equation}

and $\Delta t$ is the time step between applications of noise, usually approximated as the average length of a single gate operation.  These Gaussian distributions reproduce the $T_1$, $T_2$, and $T_\phi$ times from which they are derived, a fact we numerically verified in our implementation.

\subsection{Molecular State Preparation}
\label{sec:molecprep}
We simulate the quantum chemical state preparations within the Unitary Coupled Cluster (UCC) ansatz. Our molecular state preparation circuits are constructed from the UCC ansatz.  The defining equation of the UCC state is given by
\begin{equation}
\label{eq:expT}
\ket{\Psi^{(k)}_{CC}(\vec{\xi})} = \exp \left(\sum_k T_k(\vec{\xi})\right) \ket{\Phi_{HF}}
\end{equation}

where $\ket{\Psi_{HF}}$ is the Hartree--Fock reference state, the vector of state parameters $\vec{\xi}$ are the cluster amplitudes, and $k$ is the order of the expansion. Up to second order, these operators are given by 

\begin{equation}
\label{eq:sngexc}
T_1(\vec{\xi}) = \sum_{i_1 p_1} \xi_{i_1 p_1} (a_{i_1}^\dagger a_{p_1} - a_{p_1}^\dagger a_{i_1}) 
\end{equation}

\begin{equation}
\begin{aligned}
T_2(\vec{\xi}) = \sum_{i_1 i_2 p_1 p_2} \xi_{i_1 i_2 p_1 p_2} (a_{i_1}^\dagger a_{p_1} a_{i_2}^\dagger a_{p_2} \\
- a_{p_2}^\dagger a_{i_2} a_{p_1}^\dagger a_{i_1}) 
\end{aligned}
\end{equation}

where we index spin--orbitals occupied in the Hartree-Fock reference using $i_j$ and $p_j$ for those unoccupied.

In order to decompose Eq. \ref{eq:expT} into one- and two-qubit gates, the Trotter--Suzuki decomposition is used, defined as
\begin{equation}
\label{eq:trotter}
e^{\sum_i B_i} =  \big( \prod_i e^{B_i / \eta} \big)^\eta
\end{equation}

where the operator to be exponentiated is given as a sum of Pauli strings $B_i$, and $\eta$ is an integer called the Trotter number. The approximation becomes exact as $\eta$ approaches infinity. The effects of Trotterization order have been studied elsewhere \citep{whitfield2011,love2012,Hastings2015,babbush2015}. As the focus of the present study is the effects of decoherence, most of our quantum circuits estimate the molecular energy using a Trotter number of one.

Note that we restrict ourselves to spin--conserving excitations in the above operators (hence we are restricted to the lowest energy state corresponding to the spin of the reference state). This allows us to perform our simulations at reasonable values of the cluster amplitudes separately from the consideration of VQE optimization of these amplitudes.  However, we note that the cluster amplitudes obtained by the traditional coupled cluster calculation are expected to differ from those variationally obtained in the UCC ansatz.

To map the chemical problem to qubits from Fermions, the Jordan--Wigner (JW) or Bravyi--Kitaev (BK) mapping is used. The JW mapping uses creation and annihilation operators defined as
\begin{equation}
a_p^\dagger = ( \prod_{m < p} \sigma_m^z )\sigma_p^{+}
\end{equation}
\begin{equation}
a_p         = ( \prod_{m < p} \sigma_m^z )\sigma_p^{-}
\end{equation}
\begin{equation}
\sigma^{\pm} \equiv (\sigma^x \mp i\sigma^y )/2
\end{equation}

where $p$ is the index of a spin--orbital and $\sigma_j^\alpha$ are the Pauli matrices that act on qubit $j$. The occupation is stored \textit{locally}, i.e. each qubit stores the occupation number of a single spin--orbital. The BK transformation involves a more intricate recursive procedure that is presented elsewhere \citep{love2015}. Unlike the JW mapping, which is $O$($n$)-local, the BK mapping yields an $O$(log~$n$)-local Hamiltonian, at the expense of the spin--orbital occupancy being $O$(log~$n$)-local. It is also notable that parity storage is $O$($n$)-local and $O$(log~$n$)-local in JW and BK, respectively.

\subsection{Simulation Details}
\label{sec:simdet}

We considered the elemental diatomic molecules of the first row of the periodic table (excluding neon) at their equilibrium geometries, as well as nitrogen at several bond lengths and the isoelectronic series C$_2^{2-}$, N$_2$, and O$_2^{2+}$.

Using the PTA, we simulated quantum chemical state preparations, within the Unitary Coupled Cluster (UCC) ansatz, under three different noise scenarios: relaxation-dephasing ($T_1 = T_\phi$), pure relaxation ($T_1 \ll T_\phi$), and pure dephasing ($T_\phi \ll T_1$). Note that $T_1$ and $T_\phi$ produce the same magnitude of noise (same probability of error) for a given coherence parameter. Note also that there is an energy error and particle number error resulting from our use of only a single Trotter step in approximating the full exponential. However, our aim is to characterize the effects of noise, so all errors are reported as the difference in error between the pristine (noiseless) and noisy circuits. As noted in the introduction, Trotter errors have been analyzed in several previous studies  \citep{whitfield2011,love2012,Hastings2015,babbush2015}.

Nonetheless, we performed one set of calculations to estimate the trade-off between noise error and Trotterization error. In a noise-free circuit, using a Trotter number $\eta$=1 will be less accurate than $\eta$=2. However, the latter case produces a circuit that is twice as long, yielding a greater noise-induced error. We define $M_{Trot}$ as the coherence parameter (eqs. \ref{eq:defpxpy} \& \ref{eq:defpz}) at which the total error (Trotter error and noise error) is equal for $\eta$=1 and $\eta$=2.

In the present simulations we obtain the values of the cluster amplitudes from traditional (nonunitary) coupled cluster with single and double excitations (CCSD) from the quantum chemistry package Psi4 \citep{psi4_2012}. We used the minimal STO-3G basis set with frozen $1s^2$ core electrons and a spin-restricted reference state.  Cluster amplitudes with magnitude less than $10^{-5}$ were truncated in the experiments performed in this work. Experimental molecular geometries were taken from the NIST database \citep{nistdb}.

To obtain explicit quantum circuits, the above Fermionic cluster operators were translated into Pauli operators via the JW and BK mappings. These terms were then ordered numerically by qubit index. Finally, this sorted list of Pauli operators was translated into gate sequences of basis changes Hadamard and Yb (basis change to Y), CNOTs, and Z-rotations with the standard circuit synthesis for exponentials of tensor products of Pauli operators\citep{MikeIke}. A useful introduction to these circuits for the JW mapping can be found in Whitfield et al. \citep{whitfield2011}

\begin{figure}
\centering
\includegraphics[width=0.9\columnwidth]
{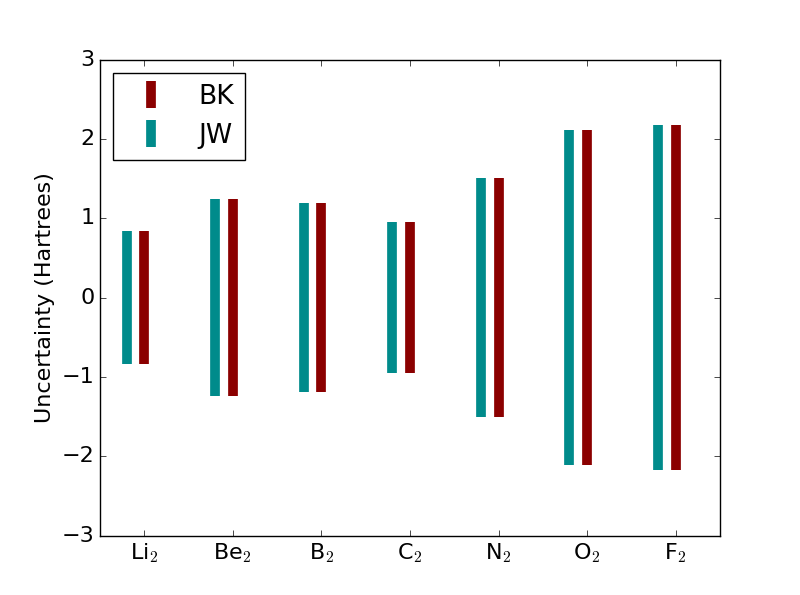}
\caption{For pristine (noise-free) circuits, the bars show one standard deviation of the chemical energy estimator (square root of the variance of the estimator). A larger standard deviation implies that a larger number of measurements is required. BK and JW denote the Bravyi--Kitaev and Jordan--Wigner mappings, respectively.}
\label{fig:nrg-pristine-F}
\end{figure}

One consideration that is especially important for VQE is the variance of the objective function estimator. Even in a pristine circuit, many measurements are required before converging on the correct answer. As an illustration, in Fig. \ref{fig:nrg-pristine-F} we show the square root of the variance for noiseless quantum circuits that simulate the first-row dimers. Hence, in addition to the mean noise-induced error, we will also report results for the noise-induced increase in variance for the chemical energy estimator, as summarized below. For VQE, an increased variance requires a larger number of quantum measurements to arrive at a particular answer. Thus trends in the variance are an important consideration, as this variance determines how many times the quantum circuit must be run.

We define operators 
\begin{equation}
\lan \cdot\ran_p = \lan \psi_p|\cdot|\psi_p\ran
\end{equation}
and
\begin{equation}
\lan \cdot\ran_n = \sum_j \lan \psi_{n,j}|\cdot|\psi_{n,j}\ran
\end{equation}

where subscripts $p$ and $n$ denote the state vectors that result from pristine (noise-free) and noisy quantum circuits, respectively, and the index $j$ runs over all noisy iterations of the simulated circuit. The estimator for the chemical energy, determined by a set of quantum measurements, is

\begin{equation}
\widehat{\lan {H} \ran} = \sum_\gamma \widehat{\lan H_\gamma\ran}
\end{equation}

where each $H_\gamma$ consists of a Pauli string, $O_\gamma$, multiplied by a constant:
\begin{equation}
H_\gamma = w_\gamma \times O_\gamma.
\end{equation}

Because variances of independent random variables are additive, we can estimate the variance in a pristine quantum circuit as
\begin{equation}
\sigma_{p}^2 = Var[\widehat{\langle H_\gamma\ran}] =
\sum_\gamma ( \langle H_\gamma^2\ran_p - \langle H_\gamma\ran_p^2 )
\end{equation}

Because $O_\gamma$ is simply a Pauli string, $\lan O_\gamma^2\ran$ is unity. Hence, we use the expression
\begin{equation}
\sigma_{p}^2 = \sum_\gamma (w_\gamma^2 - \lan H_\gamma\ran_p^2 ).
\end{equation}

Similarly, 
\begin{equation}
\sigma_{noisy}^2 = \sum_\gamma (w_\gamma^2 - \lan H_\gamma\ran_n^2 ).
\end{equation}

Finally, we define the noise-induced uncertainty or ``error width" as the quantity
\begin{equation}
\sigma_{noisy} - \sigma_p
\end{equation}

For all simulated molecules, we performed 10,000 noisy iterations of the quantum circuit.




\section{Errors in Molecular Energy}
\label{sec:enrg}

\subsection{First-Row Diatomics}

To begin studying the correlation between chemical properties and the effects of errors on quantum chemical state preparation, we first simulated a set of first-row diatomics of the periodic table. Though they are stable enough to be experimentally studied in the gas phase, several of the diatomics are unstable in air under standard laboratory conditions. However, we chose this molecular set in order to study chemical trends in susceptibility to error for the UCC ansatz.

\begin{figure}
\centering
\includegraphics[width=0.9\columnwidth]
{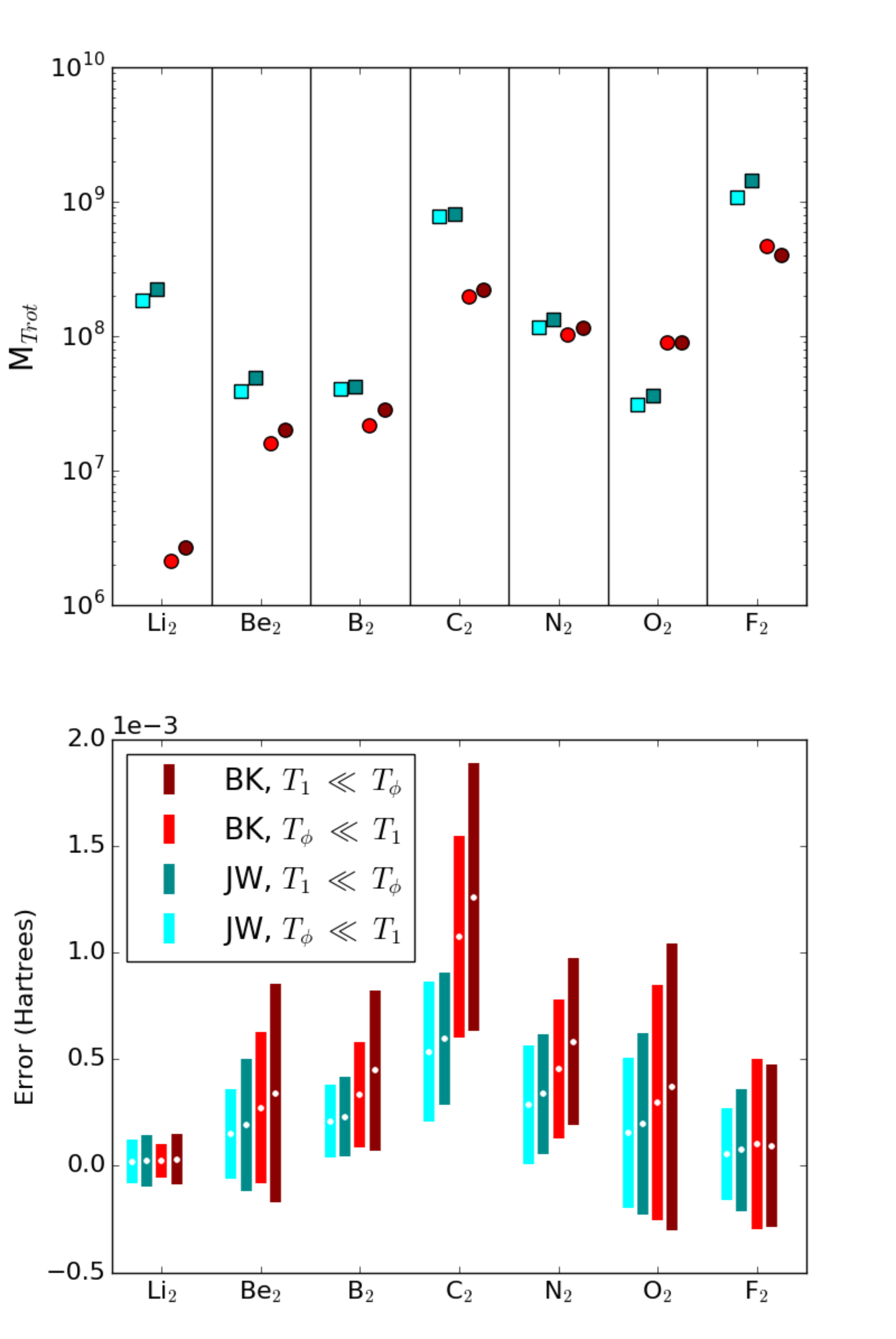}
\caption{
\textit{Top:} First row diatomic molecules. JW and BK data are plotted as squares and circles, respectively, using the color scheme of the bottom plot. The plot gives coherence parameter $M_{Trot}$, defined as the noise magnitude at which a single Trotter step produces the same error as two Trotter steps.
\textit{Bottom:} 
Mean error in the chemical energy relative to a noiseless state preparation for first row diatomics with coherence parameter $M=10^8$, in both the Jordan--Wigner and Bravyi--Kitaev encodings, for a single Trotter step. The bars represent the increase in standard deviation of the chemical energy estimator, relative to the noise-free simulation. The JW circuits exhibit less error on average than the BK mapped molecules in the cases studied. Note the multiplier corresponding to the vertical axis (1e-3).}
\label{fig:nrg-F-jwbk-m08}
\end{figure}

Fig.~\ref{fig:nrg-F-jwbk-m08} shows $M_{Trot}$ (defined in Sec. \ref{sec:simdet}) for both pure dephasing and pure relaxation, in the JW and BK mappings. $M_{Trot}$ is defined such that when $M$ = $M_{Trot}$, one cannot improve accuracy by increasing the Trotter number, because any decrease in Trotter error is counteracted by a gain in noise-induced error. Though there is no clear molecular trend, $M_{Trot}$ tends to be larger for the BK than the JW mapping. This implies that for BK there is a larger range of decoherence rates for which increasing the Trotter number is beneficial, though this analysis does not yet allow one to draw conclusions concerning which mapping produces smaller errors in noisy circuits.

The bottom plot of Fig.~\ref{fig:nrg-F-jwbk-m08} shows the distribution in chemical energy error for both JW and BK mappings, under an arbitrary coherence parameter of $M=10^8$. Recall that energy error here is defined as the difference between the measured energy and the energy of the same state preparation under noiseless conditions.  We plot results for this order of magnitude noise estimate because errors are within chemical accuracy (1 kcal/mol or $1.6\times10^{-3}$ Hartrees) for the majority of our molecular set. The JW mapping results in smaller mean error and smaller error width for a given coherence parameter, often less than half compared to the BK result. As mentioned above, error widths are related to the increase in variance of the VQE molecular energy calculation. The superior robustness of JW is especially noteworthy considering that, on average, there tend to be fewer gates in our BK circuits.

Across our full set of eighteen molecules, BK mean errors are 1.2 to 3.2 times larger for a given molecule. Error widths are up to 2.0 times larger, with the sole case of a larger JW error width occurring in the case of Li$_2$.

It is plausible that the superior robustness of JW over BK is due to the occupation numbers being stored locally. That is, a local single-qubit error in the BK mapping will affect several occupation numbers, whereas in the JW mapping it will affect just one. The corollary to this hypothesis is that occupation number errors may yield greater errors in the energy than parity errors do (see Sec. \ref{sec:molecprep}). However, further study is needed to definitively determine the source of the observed superior resistance of the JW encoding to noise, and to determine which of the two transformations is asymptotically more robust to noise as the size of the problem increases. In the remainder of this article, we use the JW mapping.

\begin{figure}
\centering
\includegraphics[width=0.9\columnwidth]
{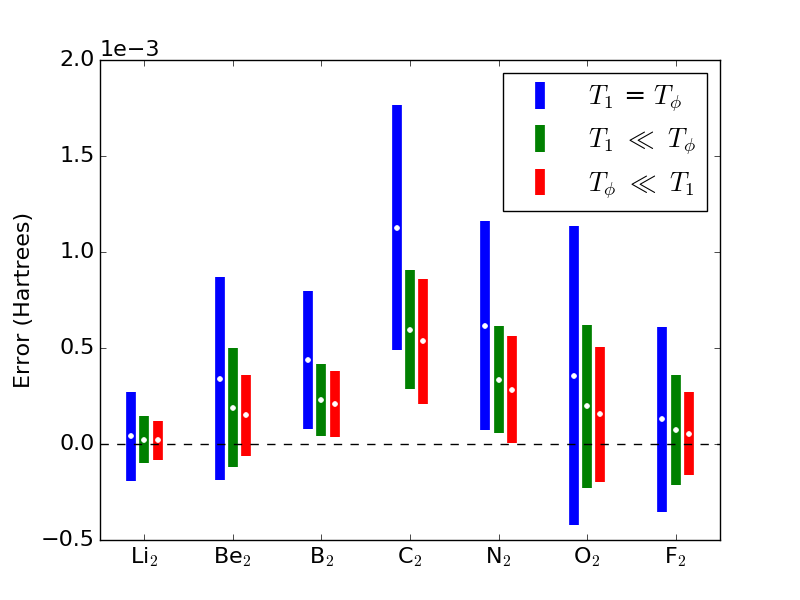}
\caption{
Mean error in the chemical energy relative to a noiseless state preparation for first row diatomics in the Jordan--Wigner encoding, under three types of noise with coherence parameter $M=10^8$.  The bars again represent the increase in standard deviation of the chemical energy estimator, relative to the noise-free simulation. In all cases, the error is more sensitive to relaxation than to dephasing. Note the multiplier corresponding to the vertical axis (1e-3).}
\label{fig:nrg-F-jw-m8}
\end{figure}

It is apparent from Fig.~\ref{fig:nrg-F-jw-m8} that pure dephasing produces consistently smaller errors than pure relaxation. This is straight-forwardly explained by noting that the larger magnitude terms in the Hamiltonian are those corresponding to expectation values of the Z operator. Consequently, the change in energy result is more sensitive to a change in reference orbital occupation number (directly related to the Z operator) because of the strength of coulombic interactions, and because particle number errors produce states outside of the manifold of physical states (see Section ~\ref{sec:enum}). On the other hand, the main effect of dephasing (related to X and Y operators in molecular energy formula) is simply to evolve the state towards a product state.  However, many of these trends in errors can become conflated with trends related to the number of gates in each quantum circuit, as longer gate sequences suffer more from decoherence.

\begin{figure}
\centering
\includegraphics[width=0.9\columnwidth]
{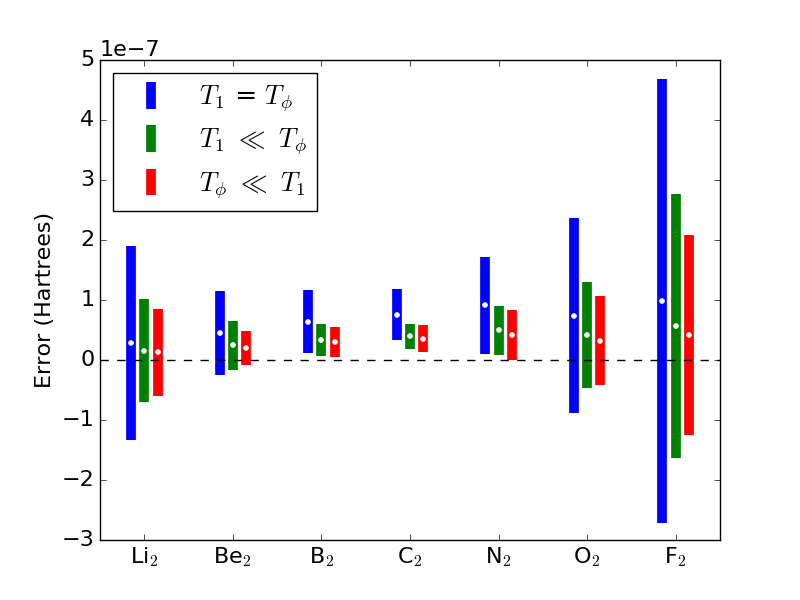}
\caption{Mean error \textit{per gate} in the chemical energy relative to a noiseless state preparation for first row diatomics in the Jordan--Wigner encoding.  The bars (error width) represent the increase in standard deviation relative to the noise-free simulation. The rough trend is for mean errors to increase with increasing nuclear charge. Error width per gate, on the other hand, increases as one moves in either direction away from carbon.
}
\label{fig:nrgPg-F-jw-m8}
\end{figure}

To help clarify the nature of these errors, we then plot energy errors per gate (equivalent to error per time step) in Fig.~\ref{fig:nrgPg-F-jw-m8}, and see a clearer trend emerge. Going to the right across the periodic table produces a roughly consistent increase in mean error. We attribute this error largely to the increase in nuclear charge, but note that many factors will combine to produce this trend, including the magnitude of terms in the Hamiltonian, the fill factor, and the distance between nuclei. The error width, on the other hand, increases as one moves in either direction away from carbon, a trend that may be attributable to the fact that C$_2$'s particle number is the best-preserved out of this set of dimers (see Sec.~\ref{sec:enum}). That is, it may be that states which are further from the manifold of physical states (physical states being those with the exactly correct electron number) have a larger error width.

\begin{figure}
\centering
\includegraphics[width=0.9\columnwidth]
{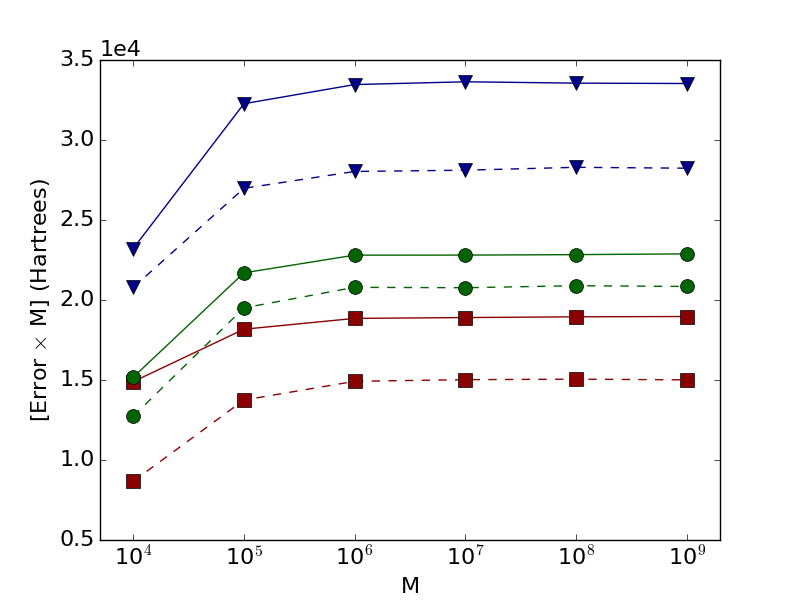}
\caption{Product of mean observed error (with respect to a noiseless preparation) and coherence parameter for Be$_2$ (red squares), B$_2$ (green circles), and N$_2$ (blue triangles) in the Jordan--Wigner mapping. Note that this plotted value is equal to the ratio of mean error to decoherence rate. Dashed lines: $T_1$ type noise only; dotted lines: $T_\phi$ type noise only. The ratio asymptotes as the noise magnitude decreases. For lower values of $M$, the final state approaches a completely mixed state, which eventually yields a constant error and therefore a decreasing error to noise ratio.}

\label{fig:nrg-BeBN-vsM}
\end{figure}

Finally, we observe that, for moderate noise magnitudes, the mean energy error to noise ratio (equal to the product of the mean energy error and the coherence parameter) is roughly constant above a certain threshold (Fig.~\ref{fig:nrg-BeBN-vsM}). We compare Be$_2$, B$_2$, and N$_2$ because their circuits contain similar total gate numbers, from around 6800 to 7600 gates. This trivial result implies that, within a large range of noise magnitudes, error decreases roughly linearly with coherence parameter, assuming uncorrelated errors. The lower values of $M$ (higher noise magnitudes) produce smaller ratios because the state approaches the completely mixed state as $M$ decreases. Since the error itself approaches a constant value as the completely mixed state is approached, the error to noise ratio decreases.

\subsection{Nitrogen Dissociation}
Next we study the dissociation of nitrogen, the primary component of Earth's atmosphere. Its dissociation is one required step of nitrogen fixation \citep{erisman2008}, necessary for producing fertilizer for the world's agricultural needs. Comparing different conformations of a particular molecule allows us to elucidate trends based on attributes such as atomic orbital overlap or nuclear separation.

\begin{figure}
\centering
\includegraphics[width=0.9\columnwidth]
{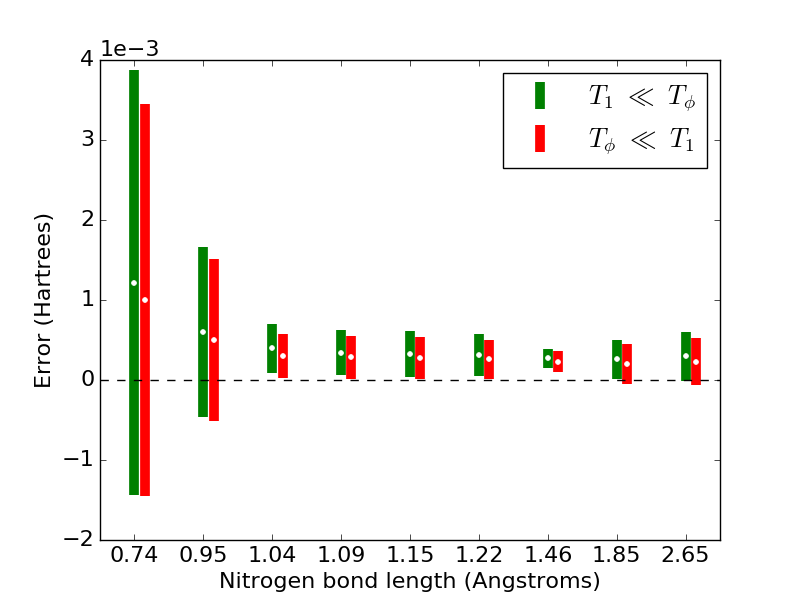}
\caption{Mean error in the energy with respect to a noiseless state preparation for several bond lengths of the N$_2$ molecule in the JW encoding. The bars enclose one standard deviation of the observed errors.  Sampling lengths are not evenly spaced. Note the multiplier corresponding to the vertical axis (1e-3).}
\label{fig:nrg-N-jw-m8}
\end{figure}
The larger error in internuclear separations of 0.95 and 0.74 \AA ~are simply a result of the larger number of gates. The varying number of gates is due in part to our using a cutoff of $10^{-5}$ for the cluster amplitudes. However, separations of 1.04 through 2.65 \AA \ have nearly identical gate counts (within 3\% of each other), which allows for more direct comparisons. We see little change in error for bond lengths greater than 1 \AA, which suggests that error behavior is similar for a broad range of bond lengths.

\begin{figure}
\centering
\includegraphics[width=0.9\columnwidth]
{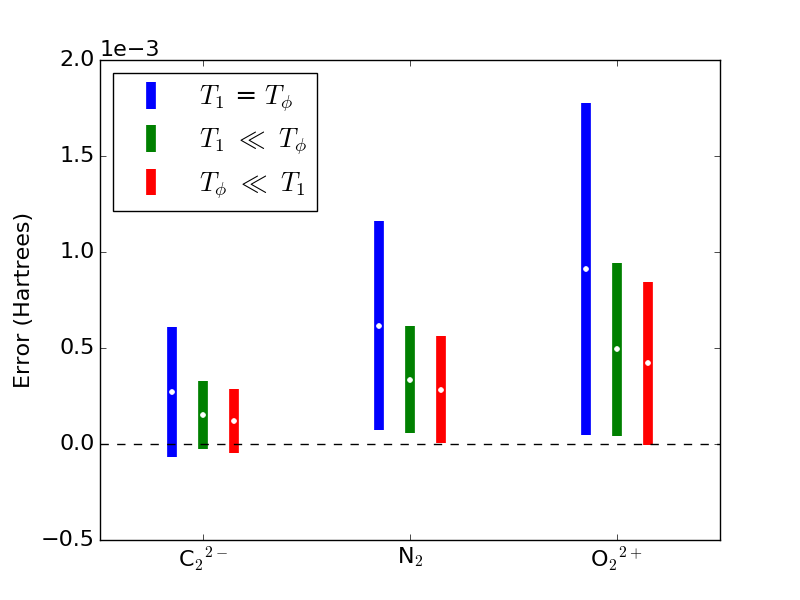}
\caption{In order to isolate nuclear charge, we plot the mean error in the energy with respect to a noisy state preparation for a series of isoelectronic species with equal bond lengths, in the JW mapping. The bars encompass one standard deviation of the observed errors. Increasing nuclear charge leads to increased error under all simulated types of noise.}
\label{fig:nrg-I-jw-m8}
\end{figure}

\subsection{Isoelectronic Species}
As a final demonstration of a simple chemical-based trend, we compare a series of compounds with an identical number of electrons but varying charge and nuclei, so-called isoelectronic species. We simulated noisy circuits for C$_2^{2-}$, N$_2$, and O$_2^{2+}$, each at the same bond length of 1.25 \AA. Because the bond length, electron count, and spin remain constant, this allows us to isolate the nuclear charge as a variable. Errors increase with increasing nuclear charge, again despite nearly identical gate counts. As before, we attribute this trend to larger magnitudes in the electronic Hamiltonian, a direct result of the increasing nuclear charges. Further connections to the previous work relating the error of gate-model quantum chemistry simulations with respect to nuclear charge \cite{babbush2015} are left for future work.

We also simulated a set of molecules in which bond length and nuclear charge were constant while the total number of electrons increased. These calculations (omitted from the article) did not produce a clear trend in molecular energy error, partly because of a large variation in the number of gates required to implement the correct cluster amplitudes. However, for all simulated molecules there is a clear trend in particle number error as electron filling varies (independent of bond length and nuclear charge), as shown in Sec.~\ref{sec:enum}.

\section{Error in Electron Number Preservation}
\label{sec:enum}

As discussed above, VQE and QPE are robust to errors assuming that we are concerned only with the final ground-state energy, not with knowing the true values of the parameters ${\xi}$ themselves. There may of course be some cases in which it is important to know the correct cluster amplitudes that correspond to the ground state. For instance, one may want to prepare the state on a different quantum device prone a different set of systematic errors, or to calculate additional state properties for which the cluster amplitudes are explicitly needed. 

Regardless of whether accurate cluster amplitude are desired, the energy calculation is useful as part of the VQE and QPE schemes only as long as the prepared state is chemically valid. In the case of the VQE approach this relates to the input state, whereas in the QPE approach this relates to the projected state at the energy measurement that could be chemically invalid as a result of Trotter or other errors.  A necessary condition for validity is that the error in the electron number operator be small. Hence, it might be considered even more important to consider the errors in the number operator than in the energy calculation, since strictly speaking errors in chemical energy do not themselves invalidate the variational principle.

In the JW mapping, an orbital in the reference state is fully occupied for qubit state $|1\rangle$ and empty for state $|0\rangle$. Hence, the total electron number operator in the second quantization formalism is written as follows:

\begin{equation}
N = \sum_i{N_i} = \sum_i{a_i^\dagger a_i}
\end{equation}

In this section we determine the error in the prepared state's electron number expectation value for the molecules listed above. Error bars in the following plots enclose one standard deviation of the error in the particle number, relative to the noise-free simulation, over the set of noisy runs (note that this definition is different from the error bars defined in the plots of chemical energy error). We also show how \qhipster ~may be useful for studying particle number errors resulting from systematic gate errors.


\subsection{Effect of Noise on Electron Number}

\begin{figure}
\centering
\includegraphics[width=0.9\columnwidth]
{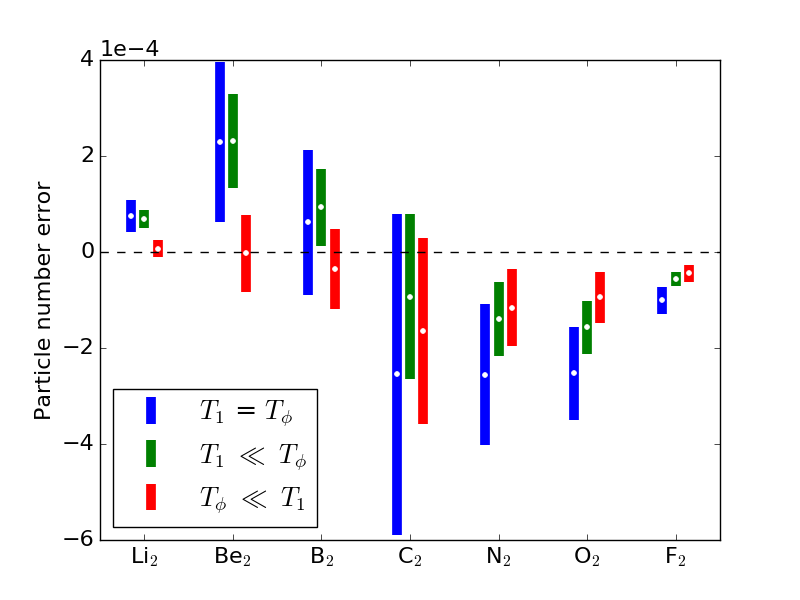}
\caption{Mean error in the electron particle number relative to a noise-free state preparation for first row diatomic species in a JW encoding.  The bars enclose one standard deviation in the observed data. Increased errors in molecules closer to the plot's center are due in part to a larger number of time steps (gates) present in the circuit. Note the multiplier corresponding to the vertical axis (1e-4).}
\label{fig:nop-F-jw-m8}
\end{figure}

\begin{figure}
\centering
\includegraphics[width=0.9\columnwidth]
{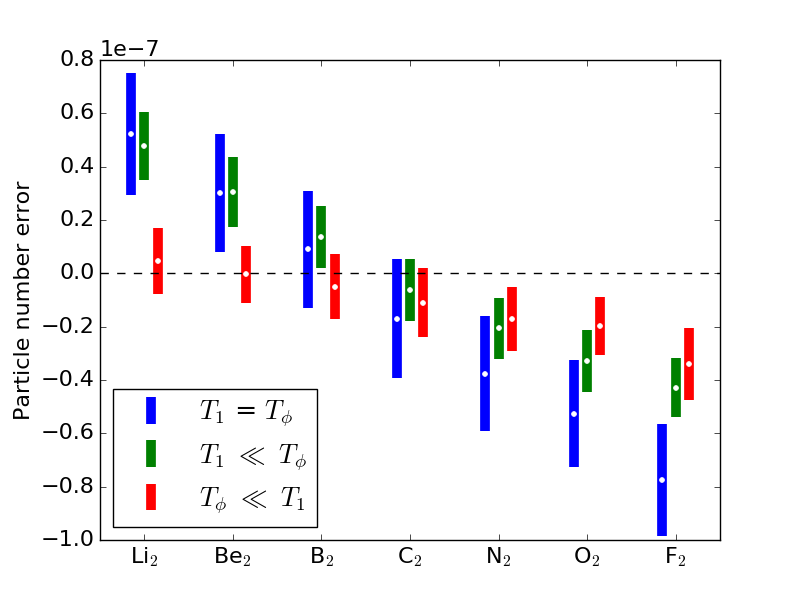}
\caption{Mean error in the electron particle number relative to a noise-free state  preparation \emph{per gate} for first row diatomic species, in a JW encoding.  The bars enclose one standard deviation in the observed data. Within our noise model, relaxation causes the particle number to approach $n/2$, where $n$ is the number of spin--orbitals. Note the multiplier corresponding to the vertical axis (1e-7).}
\label{fig:nopPg-F-jw-m8}
\end{figure}
Here we report the robustness of the electron particle number for our molecular set, in the Jordan--Wigner mapping. The simulations demonstrate that the electron particle number is significantly more robust to pure dephasing noise. This result is expected, as occupation is stored in the computational basis and all states studied in this section are close to the Hartree-Fock reference state, which is partially robust under the effects of dephasing. In other words: if no gate operations were performed on the initial reference state and pure-dephasing was present, the particle number would be preserved. It is the Hadamard and Y-basis gates that temporarily expose these Z amplitudes to error.

Insight can be gained more easily from the error per gate (Fig.~\ref{fig:nopPg-F-jw-m8}) than from the total error (Fig.~\ref{fig:nop-F-jw-m8}). The pure relaxation errors produce a clean trend, increasing based on how far the state is from the fully mixed state. This evolution towards the mixed state results from our decision to use the Pauli twirling approximation. We note that physical quantum hardware may behave asymptotically differently, depending on which quantum hardware is used. For example, qubits in ion traps \citep{Schindler2013} often are implemented in such a way that they decay to the ground state after dephasing, whereas transmon qubits \citep{martinisEcc2015} decay to the vacuum state that is outside the computational manifold. Hence our use of PTA is meant to provide estimates for error magnitudes in relaxation noise relative to dephasing noise, and we emphasize that results in real hardware may differ qualitatively from the results shown in Figures ~\ref{fig:nop-F-jw-m8} and ~\ref{fig:nopPg-F-jw-m8}.

Within our noise model, the smallest relaxation-induced errors are present in diatomics whose electron filling fractions are approximately half of the number of available spin--orbitals (B$_2$, C$_2$, N$_2$).  We attribute this to the completely mixed state being largely composed of states corresponding to half-filling, and again emphasize that this trend will depend on the asymptotic behavior of the particular quantum hardware implementation.

\begin{figure}
\centering
\includegraphics[width=0.9\columnwidth]
{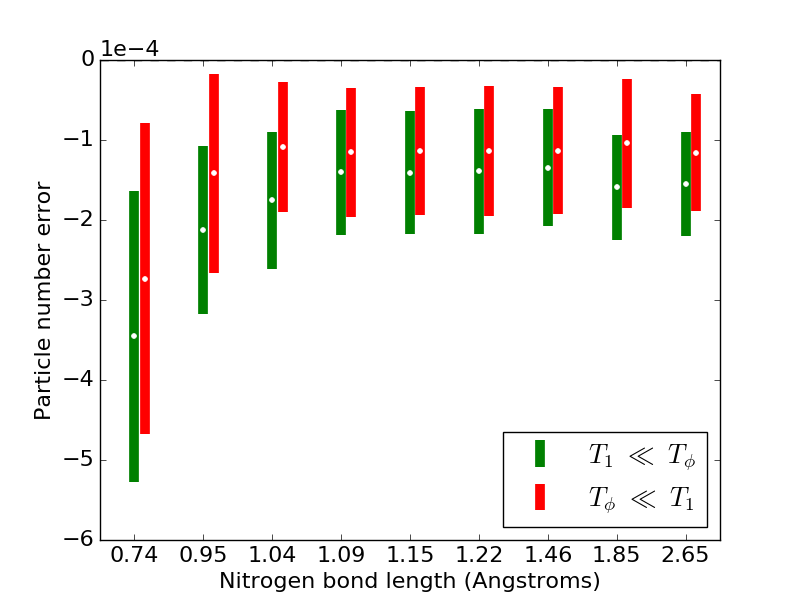}
\caption{Mean error in the electron particle number with respect to a noiseless state preparation for N$_2$ in the JW encoding at different internuclear separations. For bond distances larger than 0.95 \AA, the circuits have nearly identical gate counts (equal to the number of time steps during which noise acts on the system), which results in similar occupation number errors. }
\label{fig:nop-N-jw-m8}
\end{figure}

In nitrogen dissociation, a slightly larger error from dephasing noise implies a slightly smaller error in relaxation noise. (Again, we consider only bond lengths 1.04 through 2.65 \AA, because they have nearly identical numbers of gates.) Consider quantum circuits for bond lengths 1.46 and 1.85 \AA. The 1.46 \AA~ state's larger departure from the original Z-aligned reference state leads to qubits whose directions are less aligned with Z eigenvalues on the Bloch sphere, making them comparably less robust to dephasing noise but more robust to relaxation noise. Though dephasing noise does not appreciably affect the particle number while the qubit is not being acted upon by an exponentiation operator, it does result in larger particle number errors when the qubit is rotated into the X basis with a Hadamard, exposed to additional environmental noise, and then rotated back to the Z basis.

\subsection{Effect of Gate Errors on Electron Number}
Systematic gate errors, if present, will also affect the particle number expectation value, even in the absence of appreciable environmental errors. As mentioned above, oftentimes a key concern for a VQE type approach is the preservation of the particle number. In this section, we demonstrate the use of \qhipster~for studying simple trends in the particle number error, for up to 40 qubits. We note that, in the minimal basis of spin--orbitals (with frozen $1s^2$ electrons for heavy atoms), 40 qubits (spin--orbitals) are sufficient for simulating propane, a hydrocarbon with the chemical formula C$_3$H$_8$. For this article, we do not simulate a full quantum circuit for a specific molecule of 40 spin--orbitals, because of the long simulation times required.

As a first demonstration, we simulated two single-excitation operators, one after the other, assuming Jordan--Wigner mapping (circuit shown in reference \citep{whitfield2011}). In second quantization, this single-excitation operation is given by Eq.~\ref{eq:sngexc}.


One such circuit consists of a mix of basis-change gates (Hadamard and it Y-basis analogue, which we denote Yb), CNOT gates, and two Z-rotation gates.

For these tests, all Hadamard and Yb gates were arbitrarily given systematic over-rotation errors of $10^{-4}$ radians. Simplistic CNOT errors, where present, were represented as controlled rotations of $\pi + 10^{-3}$ radians around the X-axis. At the beginning of each simulation, the $n/4$ lowest qubits were flipped so that the initial particle number was $n/4$. All excitation magnitudes are in the range typical for quantum chemistry. Our purpose in this section is to observe the error trend, not to compare directly to gate errors in current quantum hardware. We note that our reported particle number errors are not due to Trotter errors, which our simulations showed were on the order of $10^{-14}$ or less for error-free circuits.

\begin{figure}
\centering
\includegraphics[width=0.9\columnwidth]
{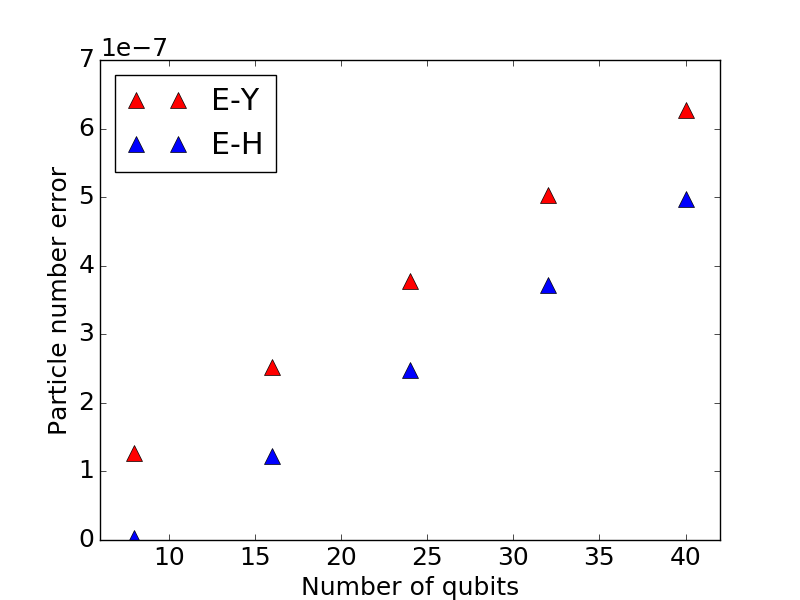}
\caption{Error in the particle number resulting from systematic single-qubit and CNOT errors, in a gate sequence representing two sequential single-excitation operators.  E-H and E-Y denote errors due to systematic over-rotations in the Hadamard and Y-basis gates, respectively. Errors increase approximately linearly with the number of qubits.}
\label{fig:numoperr-ey-eh}
\end{figure}

Figure \ref{fig:numoperr-ey-eh} shows particle number errors that result when these systematic errors are assumed to be present on all gates. The first excitation is from qubit $0$ to $n/4$, while the second is from qubit $n/4-1$ to $n-1$. All Z-rotation gates are set to 0.1 radians. Errors increase approximately linearly with the number of qubits in the simulation. For these arbitrary error magnitudes, the particle number errors from a single exponential is too small to appreciably affect the result of a VQE or QPE calculation. However, assuming multiplicative behavior and thousands of gates, these particle number errors may degrade the overall accuracy of the chemical energy calculation.

A less predictable trend appears when we vary the magnitude of the single-particle excitation, while keeping the rest of the circuit constant. We again place two single excitations in a row, varying the magnitude of the first while keeping the second fixed. There are two Z-rotations of equal magnitude associated with each of the two excitations (four Z-rotations total). The first two Z-rotations ($\theta_1$) are varied between 0 and 0.5 radians, and the second two Z-rotations ($\theta_2$) are a constant 0.1 radians. Errors were present on all basis-change gates, while systematic CNOT errors were present only on the uninterrupted set of CNOT gates immediately preceding the fourth rotation operator. Somewhat counter-intuitively, larger Z-rotations can lead to smaller particle number errors. Starting with a $\theta_1$ value of 0, the error increases slightly before declining at an increasing rate. As expected, the particle error increases with qubit number, simply because there are more total errors present in the circuit.


These circuits demonstrate the type of intuition that may be quickly gleaned by testing simple circuit trends, and one potential use of \qhipster. Alternatively, one can imagine a case in which a limited fraction of available quantum resources have high fidelities, for example if there are enough quantum resources to error-correct only some of the available qubits. A powerful classical simulator of quantum devices is useful for determining how calculation accuracies are affected by the placement of error-prone gates on a particular portion of the quantum circuit. Significant work would be needed to come to concrete conclusions concerning the optimal placement of error-prone hardware. Future studies could consider how the particle number changes after a substantial portion is already in the excited state. Finally, we note that the possibility exists to use knowledge of error trends to deliberately introduce gates into the circuit that correct the particle number, though we do not explore this possibility further here.


\begin{figure}
\centering
\includegraphics[width=0.9\columnwidth]
{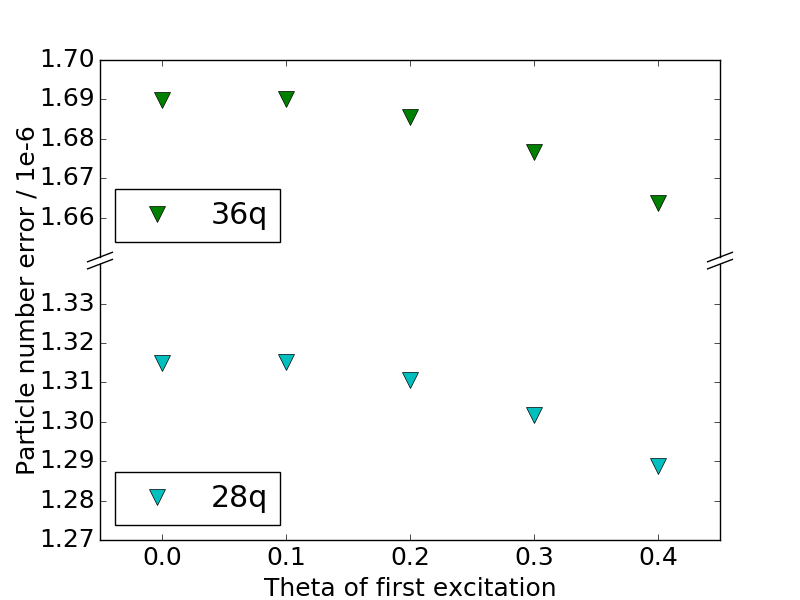}
\caption{Error in the electron particle number as a function of the excitation magnitude in the presence of systematic gate errors, for circuits of 28 and 36 qubits, in a gate sequence representing two sequential single-excitation operators. As a general trend, particle number error decreases with increasing excitation magnitude, for this set of systematic gate errors. The trend is not monotonic; of these five simulated values, the largest error occurs at $\theta = 0.1$.}
\label{fig:numoperr-vary-th}
\end{figure}

\section{High-Performance Quantum Simulator}
\label{sec:hpc}

In this section we summarize the implementation and optimization of \textit{q}H\textit{i}PSTER, our high-performance quantum simulator used to study the noisy circuits described in earlier sections. Additional details are available in ~\cite{qhipster}.

Currently, \qhipster~propagates pure states, which we use in the PTA as described. Hence, we store a $2^n\times1$ state vector instead of the $2^n\times2^n$ density matrix. As illustrated above, this still allows us to reproduce arbitrary observables of the density matrix that results from a noisy simulation.

We focus on implementing general single-qubit gates as well as two-qubit controlled gates (including CNOT gates), which are known to be universal~\cite{PhysRevA.51.1015}. Let $Q$ be a 2x2 unitary matrix that represents single qubit gate operation:
\[ 
Q=\left(
\begin{array}{cc}
q_{11} & q_{12} \\
q_{21} & q_{22} \\
\end{array} 
\right)
\]

\begin{figure*}[t!]
    \centering
    \begin{subfigure}[t]{0.18\textwidth}
        \includegraphics[width=\textwidth]{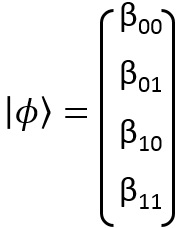}
        \caption{Quantum State}
        \label{fig:qstate}
    \end{subfigure}
    \hspace{8mm}
    \begin{subfigure}[t]{0.32\textwidth}
        \includegraphics[width=\textwidth]{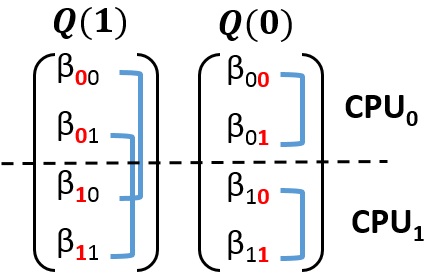}
        \caption{Single-qubit gate operation}
        \label{fig:gate}
    \end{subfigure}
    \hspace{8mm}
    \begin{subfigure}[t]{0.29\textwidth}
        \includegraphics[width=\textwidth]{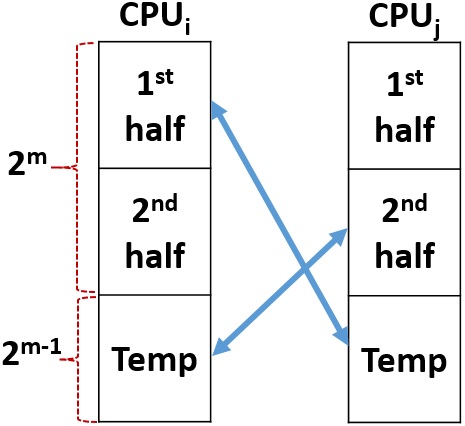}
        \caption{Distributed gate operation}
        \label{fig:gate}
    \end{subfigure}
    \caption{Example of (a) a two-qubit quantum state, (b) single-qubit gate operations, applied to qubits $0$ and $1$, respectively, and (c) an implementation of a distributed single-qubit gate operation. Communication occurs between pairs of processors, CPU$_i$ and CPU$_j$.  Processors exchange half of their states, place it into temporary storage, compute on exchanged halves, and then perform another exchange.     } \label{fig:examplequantgate}
\end{figure*}

Figure~\ref{fig:examplequantgate}a shows the vector representation of a quantum state. Each amplitude has a subscript index in the binary representation. Figure~\ref{fig:examplequantgate}b shows single-qubit gate operations on qubits 0 and 1. 
To perform a single-qubit gate on qubit $k$ of the $n$-qubit quantum register, we apply $Q$ to pairs of amplitudes whose indices differ in the $k$-th bits of their binary index:

\begin{equation}
\begin{aligned}
\beta'_{*...*0_{k}*...*} = q_{11} \cdot \beta_{*...*0_{k}*...*} + q_{12} \cdot \beta_{*...*1_{k}*...*} \\
\beta'_{*...*1_{k}*...*} = q_{21} \cdot \beta_{*...*0_{k}*...*} + q_{22} \cdot \beta_{*...*1_{k}*...*}  
\label{eq:sqgunitaryoperation}
\end{aligned}	
\end{equation}

A generalized two-qubit controlled-$Q$ gate, with  a control qubit $c$ and a target qubit $t$, works as follows: if $c$ is set to $\ket{1}$, $Q$ is applied to $t$; otherwise $t$ is left unmodified:

\begin{equation}
\begin{aligned}
\beta'_{*1_{c}*0_{t}*...*} = q_{11} \cdot \beta_{*1_{c}*0_{t}*...*} + q_{12} \cdot \beta_{*1_{c}*1_{t}*...*} \\
\beta'_{*1_{c}*1_{t}*...*} = q_{21} \cdot \beta_{*1_{c}*0_{t}*...*} + q_{22} \cdot \beta_{*1_{c}*1_{t}*...*}  
\label{eq:cqgunitaryoperation}
\end{aligned}	
\end{equation}

\subsection{Single-Node Implementation and Optimization}
\label{subsec: implandopt}

The single node implementation of a single- and two-qubit controlled gates is trivial, and directly follows from equations~\ref{eq:sqgunitaryoperation} and ~\ref{eq:cqgunitaryoperation}. Namely, we iterate over consecutive groups of amplitudes of length $2^{k+1}$, while applying $Q$ to every pair of amplitudes within the group, separated by a stride of $2^k$. Here we describe several performance optimizations.

\textit{Vectorization.} Modern Intel CPUs support SIMD (Single Instruction Multiple Data) instructions, such as AVX2~\cite{AVX2}, which can perform 4 double-precision operations simultaneously on 4 elements of the input registers. We map computation of every two pairs of complex amplitudes to 4-wide SIMD instructions. Each pair, which operates on real and imaginary parts, uses half of the SIMD register.

\textit{Multithreading.} Modern CPUs can execute  many concurrent hardware threads.   We parallelize single- and two-qubit controlled  gate operations on these threads using OpenMP 4.0~\cite{openmp13}. Groups of amplitudes are evenly divided among the threads. When there are not enough groups to utilize all available threads, we parallelize computations within a group.

\textit{Cache Blocking.} Instead of bringing the entire state from slow memory for each gate operation, we block gate operations in fast Last Level Cache (LLC). Namely, we ``fuse" groups of consecutive gates, where each gate operates on some qubit $k$,   $k < {l_c}$, where $2^l_c$ is LLC size.  We iterate over blocks of $2^{l_c}$ amplitudes of the state vector, applying each of the fused gates to this block, while the block remains resident in  LLC and therefore  benefits from the LLC's high bandwidth. 

\subsection{Multinode Implementation and Optimization}
\label{subsec: implandopt}

In our distributed implementation, a state vector of $2^n$ amplitudes ($2^{n+4}$ bytes) is distributed among $2^p$ nodes, such that each node stores a local state of $2^{n-p}$ amplitudes. Let $m=n-p$. Naturally, $2^{m+4}$ must be less than the total memory capacity of the node.  

Given single- or two-qubit  gate operation on qubit $k$, if $k < m$, the operation is fully contained within a node. If $k \geqslant m$, the first and second elements of the pair are located on two different nodes and communication is required.  We implement the communication scheme described  in~\cite{DBLP:phd/de/Trieu2010} and demonstrated in   Figure~\ref{fig:examplequantgate}c. The two nodes exchange half of their state vectors into each other's temporary storage, then compute on exchanged halves, which is followed by another pair-wise exchange.

To reduce the memory requirements of temporary storage, we divide the distributed phase into multiple steps. At each step we exchange and reserve temporary storage for only a small portion of the state vector, as opposed to the entire half, as done in reference ~\cite{DBLP:phd/de/Trieu2010} . This also allows us to overlap communication and computation in step $i$ with state exchange in steps $i-1$ and $i+2$, which helps to partially hide the overhead of communication.

\subsection{\textit{q}H\textit{i}PSTER Performance}
\label{subsec: performance}

We evaluate the performance and scalability of \textit{q}H\textit{i}PSTER on the Stampede supercomputer. Stampede~\cite{Stampede} at the Texas Advanced Computing Center (TACC)/University of Texas (\# 10 in the current TOP500 list) consists of 6,400 compute nodes, each of which is equipped with two sockets of Xeon E5-2680 and 32GB of DDR4 memory per node (16GB per socket). Each socket has 8 cores.  The nodes are connected via a Mellanox FDR 56 Gb/s InfiniBand interconnect.  Achievable memory bandwidth $B_{mem}=40$ GB/s, while achievable network bandwidth $B_{net}=5.5$ GB/s (bidirectional), per socket. 

We have used  1000 nodes (2000 sockets), the maximum available allocation. With aggregate memory capacity of 32 Tbytes across 1000 nodes, we are able to simulate a quantum system of up to 40 qubits. 

\begin{figure*}
    \centering
    \begin{subfigure}[b]{0.45\textwidth}
        \includegraphics[width=\textwidth]{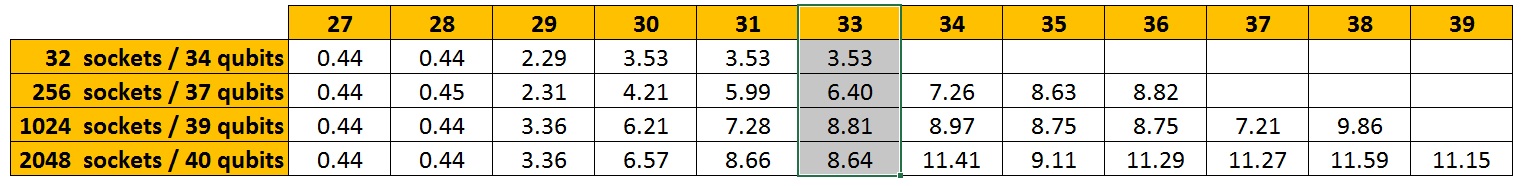}
        \caption{Weak scaling up to  40 qubits}
        \label{fig:weakscaling}
    \end{subfigure}
    \hspace{8mm}
    \begin{subfigure}[b]{0.45\textwidth}
        \includegraphics[width=\textwidth]{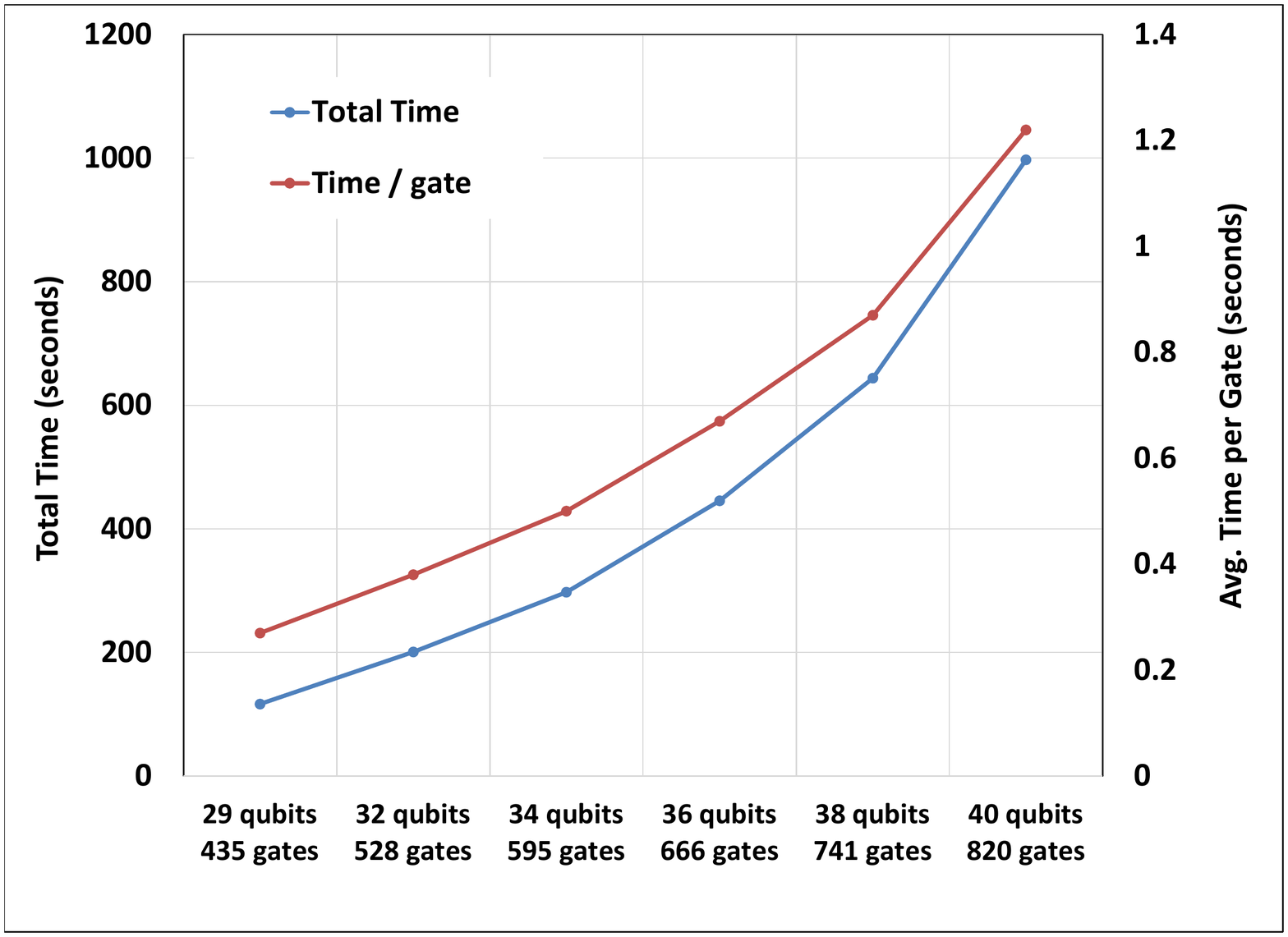}
        \caption{Quantum Fourier Transform on 40 qubits}
        \label{fig:qft}
    \end{subfigure}
    \caption{
Performance results: (a) Multinode time in seconds per single-qubit gate. Numbers next to several bars show exact measured time.  Results of gate operation qubits $0-27$ are similar to qubit $28$ and are  omitted. The large jump at 30 qubits is due to internode communication and is proportional to the ratio between network memory bandwidth. (b) Quantum Fourier Transform (QFT) up to 40 qubits. The upper curve shows total time per single QFT call. The lower curve shows average time per gate.
}
\label{fig:perfresults}
\end{figure*}

\textit{Multi-node results:} Figure~\ref{fig:weakscaling} shows time per single-qubit gate operation on multiple nodes. We report time per gate for 32, 256, 1K, and 2K sockets, which enable simulating quantum systems with 32, 37, 39, and 40 qubits, respectively.  Note this is a ``weak scaling" experiment. Specifically, we fix the local state vector to use the maximum amount of memory available on a socket. As we increase the number of qubits, we also use more sockets, and, as a result, the size of local state vector on a socket remains the same.

Gate operations applied to qubits $0-28$ require no communication for all four quantum systems, and achieve the performance of $\sim 0.44s$ per gate (shown in Figure~\ref{fig:weakscaling}, above the bar that corresponds to qubit position 28), limited only by memory bandwidth of the machine. Gates applied to higher-order qubits require communication. For the 32-node configuration we consistently see $\sim 3.53$s. This is  $\sim 8 \times$ higher than single qubit performance, and corresponds to the ratio between memory and network bandwidth of the system ($7.2= 40 / 5.5 $).

As the number of nodes increases, time per gate continues to increase. For example, a gate operation applied to qubit 39 on 1024 nodes takes 8.7 s (shown above the corresponding bar). This  is a $2.5\times$ increase, compared to applying a gate operation to qubit 33 on 32 nodes, which takes 3.53 s (also shown  above the corresponding bar). This is due to network contention and  interference with other jobs running on the system at the same time.   Controlled two-qubit gates show similar run-times and follow similar performance trends.

Note that compared to the JUMPIQCS distributed quantum simulator~\cite{DBLP:phd/de/Trieu2010}, \qhipster~ is  $3\times$ to $10\times$ faster. The advantage is due to higher memory and network bandwidth, as well as the better interconnect of Stampede system, compared to the JUMP system  of the J\"{u}lich Supercomputing
Center on which JUMPIQCS was run \footnote{In particular, Stampede uses 2-level Clos~\cite{Clos} fat tree topology which has much higher path diversity, compared to the omega interconnect of the JUMP system}.

\begin{sloppypar}
\textit{Performance of QFT.} Finally,  we report the performance of the Quantum Fourier Transform (QFT). QFT is the fundamental kernel of many quantum algorithms, such as Shor's algorithm for factoring~\cite{Shor:1997:PAP:264393.264406}, the quantum phase estimation (QPE) algorithm for estimating the eigenvalues of a unitary operator~\cite{DBLP:journals/eccc/ECCC-TR96-003}, and algorithms for the hidden subgroup problem~\cite{2004quant.ph.11037L}. Its relevance to chemistry simulation is through QPE, which is used for one of the two major algorithmic approaches for chemical quantum circuits, as discussed in the introduction.
\end{sloppypar}

Figure~\ref{fig:qft} shows the performance of QFT as the number of qubits varies from 29 to 40. Note this is also a weak scaling experiment, as the size of the local state vector per node is fixed to be $2^{29}$ complex amplitudes. We see that total QFT time varies from 116 s for 29 qubits up to 997 s for 40 qubits. On average, for 40 qubits, each QFT gate operation takes $\sim 1.22$ s, as shown in the last column of Figure~\ref{fig:qft} . 

Understanding run-time requirements of important quantum subroutines (such as QFT) is important, as it allows for estimating the simulation time of a quantum application that calls this kernel. For example, on the Stampede cluster, a single user application is limited to a maximum run-time of 24 hours. For a 40-qubit system, this would allow $\sim 86$ ($24\times3600 / 997$) calls to QFT for the total of $\sim 70,000$ quantum gates.

\section{Conclusion}
\label{sec:concl}

We have simulated noisy quantum circuits for preparing molecular electronic states in the unitary coupled cluster ansatz, in order to characterize errors in molecular energy and in electron particle number. It is necessary to study the effects of noise on the accuracy of quantum computation, because early quantum hardware will likely not be of high enough fidelity or will not contain sufficient resources to implement error correction codes. It is important to note that imperfectly error-corrected qubits will also have effective decoherence times.

For our set of eighteen molecules, we showed that the Jordan--Wigner mapping is less sensitive to noise than the Bravyi--Kitaev mapping. Isolating pure dephasing and pure relaxation noise demonstrated that relaxation noise produces larger errors than dephasing noise of the same magnitude, for both mappings. Additionally, these initial simulations suggest that there is a large range of relevant bond lengths for which error behavior will be similar.

The intuition gained from this study may be useful for guiding experimental and algorithmic choices, when implementing quantum circuits for modeling chemistry. For example, the differing sensitivities to $T_1$ and $T_\phi$ times may help in choosing between similar quantum devices that are available for use. Finally, this article has demonstrated the utility of high performance simulators such as \qhipster~ for characterizing the effects of noise and gate errors in quantum circuits of 10 to 40 qubits, especially in circuits for which analytical characterization is difficult or impossible.

\begin{acknowledgement}

The authors thank Jhonathan Romero, Alejandro Perdomo-Ortiz, Borja Peropadre, and Kostyantyn Kechedzhi for fruitful conversations, as well as Gian Giacomo Guerreschi for proofreading the manuscript. N.P.D.S. and A.A.-G. acknowledge support from the Air Force Office of Scientific Research under award FA9550-12-1-0046, the Army Research Office under award, W911NF-15-1-0256 and the Defense Security Science Engineering Fellowship managed by the Office of Naval Research under award N00014-16-1-2008. J.R.M. is a Luis W. Alvarez Fellow in Computing Sciences supported by Laboratory Directed Research and Development (LDRD) funding from Berkeley Lab, provided by the Director, Office of Science, of the U.S. Department of Energy under Contract No. DE-AC02-05CH11231. The authors also acknowledge the Texas Advanced Computing Center (TACC, http://www.tacc.utexas.edu) at the University of Texas at Austin for providing HPC resources.

\end{acknowledgement}




\bibliography{references}

\providecommand{\latin}[1]{#1}
\providecommand*\mcitethebibliography{\thebibliography}
\csname @ifundefined\endcsname{endmcitethebibliography}
  {\let\endmcitethebibliography\endthebibliography}{}
\begin{mcitethebibliography}{55}
\providecommand*\natexlab[1]{#1}
\providecommand*\mciteSetBstSublistMode[1]{}
\providecommand*\mciteSetBstMaxWidthForm[2]{}
\providecommand*\mciteBstWouldAddEndPuncttrue
  {\def\EndOfBibitem{\unskip.}}
\providecommand*\mciteBstWouldAddEndPunctfalse
  {\let\EndOfBibitem\relax}
\providecommand*\mciteSetBstMidEndSepPunct[3]{}
\providecommand*\mciteSetBstSublistLabelBeginEnd[3]{}
\providecommand*\EndOfBibitem{}
\mciteSetBstSublistMode{f}
\mciteSetBstMaxWidthForm{subitem}{(\alph{mcitesubitemcount})}
\mciteSetBstSublistLabelBeginEnd
  {\mcitemaxwidthsubitemform\space}
  {\relax}
  {\relax}

\bibitem[Montanaro(2016)]{montanaro2015}
Montanaro,~A. \emph{Npj Quantum Information} \textbf{2016}, \emph{2},
  15023\relax
\mciteBstWouldAddEndPuncttrue
\mciteSetBstMidEndSepPunct{\mcitedefaultmidpunct}
{\mcitedefaultendpunct}{\mcitedefaultseppunct}\relax
\EndOfBibitem
\bibitem[Nielsen and Chuang(2011)Nielsen, and Chuang]{MikeIke}
Nielsen,~M.~A.; Chuang,~I.~L. \emph{Quantum Computation and Quantum
  Information: 10th Anniversary Edition}, 10th ed.; Cambridge University Press:
  New York, NY, USA, 2011\relax
\mciteBstWouldAddEndPuncttrue
\mciteSetBstMidEndSepPunct{\mcitedefaultmidpunct}
{\mcitedefaultendpunct}{\mcitedefaultseppunct}\relax
\EndOfBibitem
\bibitem[Kelly \latin{et~al.}(2015)Kelly, Barends, Fowler, Megrant, Jeffrey,
  White, Sank, Mutus, Campbell, Chen, Chen, Chiaro, Dunsworth, Hoi, Neill,
  O’Malley, Quintana, Roushan, Vainsencher, Wenner, Cleland, and
  Martinis]{martinisEcc2015}
Kelly,~J.; Barends,~R.; Fowler,~A.~G.; Megrant,~A.; Jeffrey,~E.; White,~T.~C.;
  Sank,~D.; Mutus,~J.~Y.; Campbell,~B.; Chen,~Y.; Chen,~Z.; Chiaro,~B.;
  Dunsworth,~A.; Hoi,~I.-C.; Neill,~C.; O’Malley,~P. J.~J.; Quintana,~C.;
  Roushan,~P.; Vainsencher,~A.; Wenner,~J.; Cleland,~A.~N.; Martinis,~J.~M.
  \emph{Nature} \textbf{2015}, \emph{519}, 66--69\relax
\mciteBstWouldAddEndPuncttrue
\mciteSetBstMidEndSepPunct{\mcitedefaultmidpunct}
{\mcitedefaultendpunct}{\mcitedefaultseppunct}\relax
\EndOfBibitem
\bibitem[Senko \latin{et~al.}(2015)Senko, Richerme, Smith, Lee, Cohen, Retzker,
  and Monroe]{Senko2015}
Senko,~C.; Richerme,~P.; Smith,~J.; Lee,~A.; Cohen,~I.; Retzker,~A.; Monroe,~C.
  \emph{Phy. Rev. X} \textbf{2015}, \emph{5}, 1--9\relax
\mciteBstWouldAddEndPuncttrue
\mciteSetBstMidEndSepPunct{\mcitedefaultmidpunct}
{\mcitedefaultendpunct}{\mcitedefaultseppunct}\relax
\EndOfBibitem
\bibitem[Schindler \latin{et~al.}(2013)Schindler, Nigg, Monz, Barreiro,
  Martinez, Wang, Quint, Brandl, Nebendahl, Roos, Chwalla, Hennrich, and
  Blatt]{Schindler2013}
Schindler,~P.; Nigg,~D.; Monz,~T.; Barreiro,~J.~T.; Martinez,~E.; Wang,~S.~X.;
  Quint,~S.; Brandl,~M.~F.; Nebendahl,~V.; Roos,~C.~F.; Chwalla,~M.;
  Hennrich,~M.; Blatt,~R. \emph{New J. Phys.} \textbf{2013}, \emph{15},
  1--37\relax
\mciteBstWouldAddEndPuncttrue
\mciteSetBstMidEndSepPunct{\mcitedefaultmidpunct}
{\mcitedefaultendpunct}{\mcitedefaultseppunct}\relax
\EndOfBibitem
\bibitem[Rieffel and Polak(2014)Rieffel, and Polak]{RiefPolak}
Rieffel,~E.~G.; Polak,~W.~H. \emph{Quantum Computing: A Gentle Introcution};
  The MIT Press: Cambridge, MA, USA, 2014\relax
\mciteBstWouldAddEndPuncttrue
\mciteSetBstMidEndSepPunct{\mcitedefaultmidpunct}
{\mcitedefaultendpunct}{\mcitedefaultseppunct}\relax
\EndOfBibitem
\bibitem[O'Malley \latin{et~al.}(2015)O'Malley, Babbush, Kivlichan, Romero,
  McClean, Barends, Kelly, Roushan, Tranter, Ding, Campbell, Chen, Chen,
  Chiaro, Dunsworth, Fowler, Jeffrey, Megrant, Mutus, Neill, Quintana, Sank,
  Vainsencher, Wenner, White, Coveney, Love, Neven, Aspuru-Guzik, and
  Martinis]{OmalBab15}
O'Malley,~P. J.~J.; Babbush,~R.; Kivlichan,~I.~D.; Romero,~J.; McClean,~J.~R.;
  Barends,~R.; Kelly,~J.; Roushan,~P.; Tranter,~A.; Ding,~N.; Campbell,~B.;
  Chen,~Y.; Chen,~Z.; Chiaro,~B.; Dunsworth,~A.; Fowler,~A.~G.; Jeffrey,~E.;
  Megrant,~A.; Mutus,~J.~Y.; Neill,~C.; Quintana,~C.; Sank,~D.;
  Vainsencher,~A.; Wenner,~J.; White,~T.~C.; Coveney,~P.~V.; Love,~P.~J.;
  Neven,~H.; Aspuru-Guzik,~A.; Martinis,~J.~M. \emph{ArXiv e-prints}
  \textbf{2015}, arXiv:1512.06860v1\relax
\mciteBstWouldAddEndPuncttrue
\mciteSetBstMidEndSepPunct{\mcitedefaultmidpunct}
{\mcitedefaultendpunct}{\mcitedefaultseppunct}\relax
\EndOfBibitem
\bibitem[Smelyanskiy \latin{et~al.}(2016)Smelyanskiy, Sawaya, and
  Aspuru-Guzik]{qhipster}
Smelyanskiy,~M.; Sawaya,~N. P.~D.; Aspuru-Guzik,~A. \emph{ArXiv e-prints}
  \textbf{2016}, quant--ph/1601.07195\relax
\mciteBstWouldAddEndPuncttrue
\mciteSetBstMidEndSepPunct{\mcitedefaultmidpunct}
{\mcitedefaultendpunct}{\mcitedefaultseppunct}\relax
\EndOfBibitem
\bibitem[Aspuru-Guzik \latin{et~al.}(2005)Aspuru-Guzik, Dutoi, Love, and
  Head-Gordon]{aag2005}
Aspuru-Guzik,~A.; Dutoi,~A.~D.; Love,~P.~J.; Head-Gordon,~M. \emph{Science}
  \textbf{2005}, \emph{309}, 1704--1707\relax
\mciteBstWouldAddEndPuncttrue
\mciteSetBstMidEndSepPunct{\mcitedefaultmidpunct}
{\mcitedefaultendpunct}{\mcitedefaultseppunct}\relax
\EndOfBibitem
\bibitem[Abrams and Lloyd(1999)Abrams, and Lloyd]{abrams1999}
Abrams,~D.~S.; Lloyd,~S. \emph{Phys. Rev. Lett.} \textbf{1999}, \emph{83},
  5162--5165\relax
\mciteBstWouldAddEndPuncttrue
\mciteSetBstMidEndSepPunct{\mcitedefaultmidpunct}
{\mcitedefaultendpunct}{\mcitedefaultseppunct}\relax
\EndOfBibitem
\bibitem[Veis and Pittner(2010)Veis, and Pittner]{veispittner2010}
Veis,~L.; Pittner,~J. \emph{J. Chem. Phys.} \textbf{2010}, \emph{133}\relax
\mciteBstWouldAddEndPuncttrue
\mciteSetBstMidEndSepPunct{\mcitedefaultmidpunct}
{\mcitedefaultendpunct}{\mcitedefaultseppunct}\relax
\EndOfBibitem
\bibitem[Whitfield \latin{et~al.}(2011)Whitfield, Biamonte, and
  Aspuru-Guzik]{whitfield2011}
Whitfield,~J.~D.; Biamonte,~J.; Aspuru-Guzik,~A. \emph{Mol. Phys.}
  \textbf{2011}, \emph{109}, 735--750\relax
\mciteBstWouldAddEndPuncttrue
\mciteSetBstMidEndSepPunct{\mcitedefaultmidpunct}
{\mcitedefaultendpunct}{\mcitedefaultseppunct}\relax
\EndOfBibitem
\bibitem[Reiher \latin{et~al.}(2016)Reiher, Wiebe, Svore, Wecker, and
  Troyer]{reiher2016}
Reiher,~M.; Wiebe,~N.; Svore,~K.~M.; Wecker,~D.; Troyer,~M. \emph{arXiv
  preprint} \textbf{2016}, quant--ph/1605.03590v2\relax
\mciteBstWouldAddEndPuncttrue
\mciteSetBstMidEndSepPunct{\mcitedefaultmidpunct}
{\mcitedefaultendpunct}{\mcitedefaultseppunct}\relax
\EndOfBibitem
\bibitem[Peruzzo \latin{et~al.}(2015)Peruzzo, McClean, Shadbolt, Yung, Zhou,
  Love, Aspuru-Guzik, and O’Brien]{peruzzomcclean2014}
Peruzzo,~A.; McClean,~J.; Shadbolt,~P.; Yung,~M.-H.; Zhou,~X.-Q.; Love,~P.~J.;
  Aspuru-Guzik,~A.; O’Brien,~J.~L. \emph{Nat. Commun.} \textbf{2015},
  \emph{5}, 4213\relax
\mciteBstWouldAddEndPuncttrue
\mciteSetBstMidEndSepPunct{\mcitedefaultmidpunct}
{\mcitedefaultendpunct}{\mcitedefaultseppunct}\relax
\EndOfBibitem
\bibitem[McClean \latin{et~al.}(2016)McClean, Romero, Babbush, and
  Aspuru-Guzik]{mcclean2015}
McClean,~J.~R.; Romero,~J.; Babbush,~R.; Aspuru-Guzik,~A. \emph{New J. Phys.}
  \textbf{2016}, \emph{18}, 023023\relax
\mciteBstWouldAddEndPuncttrue
\mciteSetBstMidEndSepPunct{\mcitedefaultmidpunct}
{\mcitedefaultendpunct}{\mcitedefaultseppunct}\relax
\EndOfBibitem
\bibitem[Wecker \latin{et~al.}(2015)Wecker, Hastings, and
  Troyer]{wecker2015_vqe}
Wecker,~D.; Hastings,~M.~B.; Troyer,~M. \emph{Phys. Rev. A} \textbf{2015},
  \emph{92}, 042303\relax
\mciteBstWouldAddEndPuncttrue
\mciteSetBstMidEndSepPunct{\mcitedefaultmidpunct}
{\mcitedefaultendpunct}{\mcitedefaultseppunct}\relax
\EndOfBibitem
\bibitem[Barrett \latin{et~al.}(2013)Barrett, Hammerer, Harrison, Northup, and
  Osborne]{barrett2013}
Barrett,~S.; Hammerer,~K.; Harrison,~S.; Northup,~T.~E.; Osborne,~T.~J.
  \emph{Phys. Rev. Lett.} \textbf{2013}, \emph{110}, 090501\relax
\mciteBstWouldAddEndPuncttrue
\mciteSetBstMidEndSepPunct{\mcitedefaultmidpunct}
{\mcitedefaultendpunct}{\mcitedefaultseppunct}\relax
\EndOfBibitem
\bibitem[Eichler \latin{et~al.}(2015)Eichler, Mlynek, Butscher, Kurpiers,
  Hammerer, Osborne, and Wallraff]{eichler2015}
Eichler,~C.; Mlynek,~J.; Butscher,~J.; Kurpiers,~P.; Hammerer,~K.;
  Osborne,~T.~J.; Wallraff,~A. \emph{Phys. Rev. X} \textbf{2015}, \emph{5},
  041044\relax
\mciteBstWouldAddEndPuncttrue
\mciteSetBstMidEndSepPunct{\mcitedefaultmidpunct}
{\mcitedefaultendpunct}{\mcitedefaultseppunct}\relax
\EndOfBibitem
\bibitem[Taube and Bartlett(2006)Taube, and Bartlett]{ucc2006}
Taube,~A.~G.; Bartlett,~R.~J. \emph{Int. J. Quant. Chem.} \textbf{2006},
  \emph{106}, 3393--3401\relax
\mciteBstWouldAddEndPuncttrue
\mciteSetBstMidEndSepPunct{\mcitedefaultmidpunct}
{\mcitedefaultendpunct}{\mcitedefaultseppunct}\relax
\EndOfBibitem
\bibitem[Bartlett and Musia\l{}(2007)Bartlett, and Musia\l{}]{ucc2007}
Bartlett,~R.~J.; Musia\l{},~M. \emph{Rev. Mod. Phys.} \textbf{2007}, \emph{79},
  291--352\relax
\mciteBstWouldAddEndPuncttrue
\mciteSetBstMidEndSepPunct{\mcitedefaultmidpunct}
{\mcitedefaultendpunct}{\mcitedefaultseppunct}\relax
\EndOfBibitem
\bibitem[Wang \latin{et~al.}(2008)Wang, Kais, Aspuru-Guzik, and
  Hoffmann]{Wang:2008}
Wang,~H.; Kais,~S.; Aspuru-Guzik,~A.; Hoffmann,~M.~R. \emph{Phys. Chem. Chem.
  Phys.} \textbf{2008}, \emph{10}, 5388--5393\relax
\mciteBstWouldAddEndPuncttrue
\mciteSetBstMidEndSepPunct{\mcitedefaultmidpunct}
{\mcitedefaultendpunct}{\mcitedefaultseppunct}\relax
\EndOfBibitem
\bibitem[McClean \latin{et~al.}(2014)McClean, Babbush, Love, and
  Aspuru-Guzik]{McClean2014b}
McClean,~J.~R.; Babbush,~R.; Love,~P.~J.; Aspuru-Guzik,~A. \emph{J. Phys. Chem.
  Lett.} \textbf{2014}, \emph{5}, 4368--4380\relax
\mciteBstWouldAddEndPuncttrue
\mciteSetBstMidEndSepPunct{\mcitedefaultmidpunct}
{\mcitedefaultendpunct}{\mcitedefaultseppunct}\relax
\EndOfBibitem
\bibitem[Jordan and Wigner()Jordan, and Wigner]{jw1928}
Jordan,~P.; Wigner,~E. \emph{Zeitschrift f{\"u}r Physik} \emph{47},
  631--651\relax
\mciteBstWouldAddEndPuncttrue
\mciteSetBstMidEndSepPunct{\mcitedefaultmidpunct}
{\mcitedefaultendpunct}{\mcitedefaultseppunct}\relax
\EndOfBibitem
\bibitem[Bravyi and Kitaev(2002)Bravyi, and Kitaev]{bk2002}
Bravyi,~S.~B.; Kitaev,~A.~Y. \emph{Ann. Phys.} \textbf{2002}, \emph{298}, 210
  -- 226\relax
\mciteBstWouldAddEndPuncttrue
\mciteSetBstMidEndSepPunct{\mcitedefaultmidpunct}
{\mcitedefaultendpunct}{\mcitedefaultseppunct}\relax
\EndOfBibitem
\bibitem[Seeley \latin{et~al.}(2012)Seeley, Richard, and Love]{love2012}
Seeley,~J.~T.; Richard,~M.~J.; Love,~P.~J. \emph{J. Chem. Phys.} \textbf{2012},
  \emph{137}, 224109\relax
\mciteBstWouldAddEndPuncttrue
\mciteSetBstMidEndSepPunct{\mcitedefaultmidpunct}
{\mcitedefaultendpunct}{\mcitedefaultseppunct}\relax
\EndOfBibitem
\bibitem[Hastings \latin{et~al.}(2015)Hastings, Wecker, Bauer, and
  Troyer]{Hastings2015}
Hastings,~M.~B.; Wecker,~D.; Bauer,~B.; Troyer,~M. \emph{Quantum Info. Comput.}
  \textbf{2015}, \emph{15}, 1--21\relax
\mciteBstWouldAddEndPuncttrue
\mciteSetBstMidEndSepPunct{\mcitedefaultmidpunct}
{\mcitedefaultendpunct}{\mcitedefaultseppunct}\relax
\EndOfBibitem
\bibitem[Babbush \latin{et~al.}(2015)Babbush, McClean, Wecker, Aspuru-Guzik,
  and Wiebe]{babbush2015}
Babbush,~R.; McClean,~J.; Wecker,~D.; Aspuru-Guzik,~A.; Wiebe,~N. \emph{Phys.
  Rev. A} \textbf{2015}, \emph{91}, 022311\relax
\mciteBstWouldAddEndPuncttrue
\mciteSetBstMidEndSepPunct{\mcitedefaultmidpunct}
{\mcitedefaultendpunct}{\mcitedefaultseppunct}\relax
\EndOfBibitem
\bibitem[Makri and Makarov(1995)Makri, and Makarov]{quapi1}
Makri,~N.; Makarov,~D.~E. \emph{J. Chem. Phys.} \textbf{1995}, \emph{102},
  4600--4610\relax
\mciteBstWouldAddEndPuncttrue
\mciteSetBstMidEndSepPunct{\mcitedefaultmidpunct}
{\mcitedefaultendpunct}{\mcitedefaultseppunct}\relax
\EndOfBibitem
\bibitem[Makri and Makarov(1995)Makri, and Makarov]{quapi2}
Makri,~N.; Makarov,~D.~E. \emph{J. Chem. Phys.} \textbf{1995}, \emph{102},
  4611--4618\relax
\mciteBstWouldAddEndPuncttrue
\mciteSetBstMidEndSepPunct{\mcitedefaultmidpunct}
{\mcitedefaultendpunct}{\mcitedefaultseppunct}\relax
\EndOfBibitem
\bibitem[Redfield(1965)]{redfield1965}
Redfield,~A.~G. In \emph{Adv. Magn. Reson.}; Waugh,~J.~S., Ed.; Advances in
  Magnetic and Optical Resonance; Academic Press, 1965; Vol.~1; pp 1 --
  32\relax
\mciteBstWouldAddEndPuncttrue
\mciteSetBstMidEndSepPunct{\mcitedefaultmidpunct}
{\mcitedefaultendpunct}{\mcitedefaultseppunct}\relax
\EndOfBibitem
\bibitem[Plenio and Knight(1998)Plenio, and Knight]{Plenio98}
Plenio,~M.~B.; Knight,~P.~L. \emph{Rev. Mod. Phys.} \textbf{1998}, \emph{70},
  101--144\relax
\mciteBstWouldAddEndPuncttrue
\mciteSetBstMidEndSepPunct{\mcitedefaultmidpunct}
{\mcitedefaultendpunct}{\mcitedefaultseppunct}\relax
\EndOfBibitem
\bibitem[Fox(1978)]{Fox1978}
Fox,~R.~F. \emph{Phys. Rep.} \textbf{1978}, \emph{48}, 179 -- 283\relax
\mciteBstWouldAddEndPuncttrue
\mciteSetBstMidEndSepPunct{\mcitedefaultmidpunct}
{\mcitedefaultendpunct}{\mcitedefaultseppunct}\relax
\EndOfBibitem
\bibitem[Aharonov \latin{et~al.}(2006)Aharonov, Kitaev, and
  Preskill]{preskill2006}
Aharonov,~D.; Kitaev,~A.; Preskill,~J. \emph{Phys. Rev. Lett.} \textbf{2006},
  \emph{96}, 050504\relax
\mciteBstWouldAddEndPuncttrue
\mciteSetBstMidEndSepPunct{\mcitedefaultmidpunct}
{\mcitedefaultendpunct}{\mcitedefaultseppunct}\relax
\EndOfBibitem
\bibitem[Monz \latin{et~al.}(2011)Monz, Schindler, Barreiro, Chwalla, Nigg,
  Coish, Harlander, H{\"{a}}nsel, Hennrich, and Blatt]{Monz11}
Monz,~T.; Schindler,~P.; Barreiro,~J.~T.; Chwalla,~M.; Nigg,~D.; Coish,~W.~a.;
  Harlander,~M.; H{\"{a}}nsel,~W.; Hennrich,~M.; Blatt,~R. \emph{Phys. Rev.
  Lett.} \textbf{2011}, \emph{106}, 130506\relax
\mciteBstWouldAddEndPuncttrue
\mciteSetBstMidEndSepPunct{\mcitedefaultmidpunct}
{\mcitedefaultendpunct}{\mcitedefaultseppunct}\relax
\EndOfBibitem
\bibitem[Kitaev(1997)]{solovaykitaev}
Kitaev,~A.~Y. \emph{Russ. Math. Surv.} \textbf{1997}, \emph{52}, 1191\relax
\mciteBstWouldAddEndPuncttrue
\mciteSetBstMidEndSepPunct{\mcitedefaultmidpunct}
{\mcitedefaultendpunct}{\mcitedefaultseppunct}\relax
\EndOfBibitem
\bibitem[Jacobs and Knight(1998)Jacobs, and Knight]{jacobs1998}
Jacobs,~K.; Knight,~P.~L. \emph{Phys. Rev. A} \textbf{1998}, \emph{57},
  2301--2310\relax
\mciteBstWouldAddEndPuncttrue
\mciteSetBstMidEndSepPunct{\mcitedefaultmidpunct}
{\mcitedefaultendpunct}{\mcitedefaultseppunct}\relax
\EndOfBibitem
\bibitem[Bassi(2003)]{bassi2003}
Bassi,~A. \emph{Phys. Rev. A} \textbf{2003}, \emph{67}, 062101\relax
\mciteBstWouldAddEndPuncttrue
\mciteSetBstMidEndSepPunct{\mcitedefaultmidpunct}
{\mcitedefaultendpunct}{\mcitedefaultseppunct}\relax
\EndOfBibitem
\bibitem[Bassi and Deckert(2008)Bassi, and Deckert]{bassi08}
Bassi,~A.; Deckert,~D.-A. \emph{Phys. Rev. A} \textbf{2008}, \emph{77},
  032323\relax
\mciteBstWouldAddEndPuncttrue
\mciteSetBstMidEndSepPunct{\mcitedefaultmidpunct}
{\mcitedefaultendpunct}{\mcitedefaultseppunct}\relax
\EndOfBibitem
\bibitem[Silva \latin{et~al.}(2008)Silva, Magesan, Kribs, and
  Emerson]{silva2008}
Silva,~M.; Magesan,~E.; Kribs,~D.~W.; Emerson,~J. \emph{Phys. Rev. A}
  \textbf{2008}, \emph{78}, 012347\relax
\mciteBstWouldAddEndPuncttrue
\mciteSetBstMidEndSepPunct{\mcitedefaultmidpunct}
{\mcitedefaultendpunct}{\mcitedefaultseppunct}\relax
\EndOfBibitem
\bibitem[Geller and Zhou(2013)Geller, and Zhou]{geller2013}
Geller,~M.~R.; Zhou,~Z. \emph{Phys. Rev. A} \textbf{2013}, \emph{88},
  012314\relax
\mciteBstWouldAddEndPuncttrue
\mciteSetBstMidEndSepPunct{\mcitedefaultmidpunct}
{\mcitedefaultendpunct}{\mcitedefaultseppunct}\relax
\EndOfBibitem
\bibitem[Tomita and Svore(2014)Tomita, and Svore]{tomita2014}
Tomita,~Y.; Svore,~K.~M. \emph{Phys. Rev. A} \textbf{2014}, \emph{90},
  062320\relax
\mciteBstWouldAddEndPuncttrue
\mciteSetBstMidEndSepPunct{\mcitedefaultmidpunct}
{\mcitedefaultendpunct}{\mcitedefaultseppunct}\relax
\EndOfBibitem
\bibitem[Trieu(2010)]{DBLP:phd/de/Trieu2010}
Trieu,~D.~B. Large-scale simulations of error prone quantum computation
  devices. Ph.D.\ thesis, University of Wuppertal, 2010\relax
\mciteBstWouldAddEndPuncttrue
\mciteSetBstMidEndSepPunct{\mcitedefaultmidpunct}
{\mcitedefaultendpunct}{\mcitedefaultseppunct}\relax
\EndOfBibitem
\bibitem[Tranter \latin{et~al.}(2015)Tranter, Sofia, Seeley, Kaicher, McClean,
  Babbush, Coveney, Mintert, Wilhelm, and Love]{love2015}
Tranter,~A.; Sofia,~S.; Seeley,~J.; Kaicher,~M.; McClean,~J.; Babbush,~R.;
  Coveney,~P.~V.; Mintert,~F.; Wilhelm,~F.; Love,~P.~J. \emph{Int. J. Quant.
  Chem.} \textbf{2015}, \emph{115}, 1431--1441\relax
\mciteBstWouldAddEndPuncttrue
\mciteSetBstMidEndSepPunct{\mcitedefaultmidpunct}
{\mcitedefaultendpunct}{\mcitedefaultseppunct}\relax
\EndOfBibitem
\bibitem[Turney \latin{et~al.}(2012)Turney, Simmonett, Parrish, Hohenstein,
  Evangelista, Fermann, Mintz, Burns, Wilke, Abrams, Russ, Leininger, Janssen,
  Seidl, Allen, Schaefer, King, Valeev, Sherrill, and Crawford]{psi4_2012}
Turney,~J.~M.; Simmonett,~A.~C.; Parrish,~R.~M.; Hohenstein,~E.~G.;
  Evangelista,~F.~A.; Fermann,~J.~T.; Mintz,~B.~J.; Burns,~L.~A.; Wilke,~J.~J.;
  Abrams,~M.~L.; Russ,~N.~J.; Leininger,~M.~L.; Janssen,~C.~L.; Seidl,~E.~T.;
  Allen,~W.~D.; Schaefer,~H.~F.; King,~R.~A.; Valeev,~E.~F.; Sherrill,~C.~D.;
  Crawford,~T.~D. \emph{WIREs Comput. Mol. Sci.} \textbf{2012}, \emph{2},
  556--565\relax
\mciteBstWouldAddEndPuncttrue
\mciteSetBstMidEndSepPunct{\mcitedefaultmidpunct}
{\mcitedefaultendpunct}{\mcitedefaultseppunct}\relax
\EndOfBibitem
\bibitem[nis(2015)]{nistdb}
{NIST Computational Chemistry Comparison and Benchmark Database, NIST Standard
  Reference Database Number 101 Release 17b}. 2015;
  \url{http://cccbdb.nist.gov/}\relax
\mciteBstWouldAddEndPuncttrue
\mciteSetBstMidEndSepPunct{\mcitedefaultmidpunct}
{\mcitedefaultendpunct}{\mcitedefaultseppunct}\relax
\EndOfBibitem
\bibitem[Erisman \latin{et~al.}(2008)Erisman, Sutton, Galloway, Klimont, and
  Winiwarter]{erisman2008}
Erisman,~J.~W.; Sutton,~M.~A.; Galloway,~J.; Klimont,~Z.; Winiwarter,~W.
  \emph{Nat. Geosci.} \textbf{2008}, \emph{1}, 636\relax
\mciteBstWouldAddEndPuncttrue
\mciteSetBstMidEndSepPunct{\mcitedefaultmidpunct}
{\mcitedefaultendpunct}{\mcitedefaultseppunct}\relax
\EndOfBibitem
\bibitem[DiVincenzo(1995)]{PhysRevA.51.1015}
DiVincenzo,~D.~P. \emph{Phys. Rev. A} \textbf{1995}, \emph{51},
  1015--1022\relax
\mciteBstWouldAddEndPuncttrue
\mciteSetBstMidEndSepPunct{\mcitedefaultmidpunct}
{\mcitedefaultendpunct}{\mcitedefaultseppunct}\relax
\EndOfBibitem
\bibitem[AVX()]{AVX2}
{New AVX2 instruction descriptions available.} {software.intel.com}, Retrieved:
  2012-01-17\relax
\mciteBstWouldAddEndPuncttrue
\mciteSetBstMidEndSepPunct{\mcitedefaultmidpunct}
{\mcitedefaultendpunct}{\mcitedefaultseppunct}\relax
\EndOfBibitem
\bibitem[{OpenMP Architecture Review Board}(2013)]{openmp13}
{OpenMP Architecture Review Board}, {OpenMP} Application Program Interface
  Version 3.0. 2013;
  \url{http://www.openmp.org/mp-documents/OpenMP4.0.0.pdf}\relax
\mciteBstWouldAddEndPuncttrue
\mciteSetBstMidEndSepPunct{\mcitedefaultmidpunct}
{\mcitedefaultendpunct}{\mcitedefaultseppunct}\relax
\EndOfBibitem
\bibitem[Sta()]{Stampede}
{Texas Advanced Computing Center (TACC)}.
  {https://www.tacc.utexas.edu/stampede/}, Retrieved: 2014-12-01\relax
\mciteBstWouldAddEndPuncttrue
\mciteSetBstMidEndSepPunct{\mcitedefaultmidpunct}
{\mcitedefaultendpunct}{\mcitedefaultseppunct}\relax
\EndOfBibitem
\bibitem[Clos(1953)]{Clos}
Clos,~C. \emph{Bell Syst. Tech. J.} \textbf{1953}, \emph{32}, 406--424\relax
\mciteBstWouldAddEndPuncttrue
\mciteSetBstMidEndSepPunct{\mcitedefaultmidpunct}
{\mcitedefaultendpunct}{\mcitedefaultseppunct}\relax
\EndOfBibitem
\bibitem[Shor(1997)]{Shor:1997:PAP:264393.264406}
Shor,~P.~W. \emph{SIAM J. Comput.} \textbf{1997}, \emph{26}, 1484--1509\relax
\mciteBstWouldAddEndPuncttrue
\mciteSetBstMidEndSepPunct{\mcitedefaultmidpunct}
{\mcitedefaultendpunct}{\mcitedefaultseppunct}\relax
\EndOfBibitem
\bibitem[Kitaev(1996)]{DBLP:journals/eccc/ECCC-TR96-003}
Kitaev,~A. \emph{Electronic Colloquium on Computational Complexity {(ECCC)}}
  \textbf{1996}, \emph{3}\relax
\mciteBstWouldAddEndPuncttrue
\mciteSetBstMidEndSepPunct{\mcitedefaultmidpunct}
{\mcitedefaultendpunct}{\mcitedefaultseppunct}\relax
\EndOfBibitem
\bibitem[{Lomont}(2004)]{2004quant.ph.11037L}
{Lomont},~C. \emph{ArXiv e-prints} \textbf{2004}, quant--ph/0411037v1\relax
\mciteBstWouldAddEndPuncttrue
\mciteSetBstMidEndSepPunct{\mcitedefaultmidpunct}
{\mcitedefaultendpunct}{\mcitedefaultseppunct}\relax
\EndOfBibitem
\end{mcitethebibliography}

\end{document}